\documentclass[aps,prl,reprint,groupedaddress,showpacs,amsmath,amssymb,twocolumn,floatfix]{revtex4-1}
\usepackage{graphicx}
\usepackage{bm}
\usepackage{color}
\usepackage[normalem]{ulem}

\usepackage{amsmath}

\newcommand{\Rxx}{$R_{xx}$}
\newcommand{\Rxy}{$R_{xy}$}

\newcommand{\muu}{$\mu_{tTLG}$}

\newcommand{\WSe}{WSe$_2$}

\usepackage{pifont}
\newcommand{\cmark}{\ding{51}}
\newcommand{\xmark}{\ding{55}}

\newcommand{\equref}[1]{Eq.~(\ref{#1})}
\usepackage{bbm}
\newcommand{\pdagger}{{\phantom{\dagger}}}
\usepackage{enumitem}   
\usepackage{tabularx}

\usepackage[colorlinks=true]{hyperref}  
\hypersetup{
    bookmarks=true,         
    unicode=false,          
    pdftoolbar=true,        
    pdfmenubar=true,        
    pdffitwindow=false,     
    pdfstartview={FitH},    
    pdftitle={twist angle dependence in twisted trilayer graphene},    
    pdfauthor={Siriviboon et al.},     
    pdfsubject={},   
    pdfcreator={},   
    pdfproducer={}, 
    pdfkeywords={} {} {}, 
    pdfnewwindow=true,      
    colorlinks=true,       
    linkcolor=magenta, 
    citecolor=blue,        
    filecolor=magenta,      
    urlcolor=blue           
} 
\usepackage{cleveref}
\begin{document}

\title{A new flavor of correlation and superconductivity in small twist-angle trilayer graphene}

\author{Phum Siriviboon$^{1}$$^{\ast}$}
\author{Jiang-Xiazi Lin$^{1}$$^{\ast}$}
\author{Xiaoxue Liu$^{1}$$^{\ast}$}
\author{Harley D. Scammell$^{2}$}
\author{Song Liu$^{3}$}
\author{Daniel Rhodes$^{3}$}
\author{K. Watanabe$^{4}$}
\author{T. Taniguchi$^{5}$}
\author{James Hone$^{3}$}
\author{Mathias S. Scheurer$^{6}$}
\author{J.I.A. Li$^{1}$}
\email{jia\_li@brown.edu}

\affiliation{$^{1}$Department of Physics, Brown University, Providence, RI 02912, USA}
\affiliation{$^{2}$School of Physics, the University of New South Wales, Sydney, NSW, 2052
Australia}
\affiliation{$^{3}$Department of Mechanical Engineering, Columbia University, New York, NY 10027, USA}
\affiliation{$^{4}$Research Center for Functional Materials, National Institute for Materials Science, 1-1 Namiki, Tsukuba 305-0044, Japan}
\affiliation{$^{5}$International Center for Materials Nanoarchitectonics,
National Institute for Materials Science,  1-1 Namiki, Tsukuba 305-0044, Japan}
\affiliation{$^{6}$ Institute for Theoretical Physics, University of Innsbruck, Innsbruck, A-6020, Austria}

\affiliation{$^{\ast}$These authors contributed equally in this work.}
\date{\today}

\maketitle

\textbf{When layers of graphene are rotationally misaligned by the magic angle, the moir\'e superlattice features extremely flat bands. Due to the enhanced density of states, the Coulomb interaction induces a variety of instabilities. The most prominent occur at integer filling and are therefore commonly attributed to spontaneous polarization of the moir\'e unit cell’s `flavor' degrees of freedom---spin, valley, and the flat-band degeneracy. 
As the dominant member of the hierarchy, these correlated states are thought to crucially determine further instabilities at lower energy scales, such as superconductivity and weaker incompressible states at fractional filling.
In this work, we examine the behavior of twisted trilayer graphene in a window of twist angle around $1.3^{\circ}$, well below the expected magic angle of $1.55^{\circ}$. 
In this small twist angle regime, we find surprisingly narrow bands, which are populated with both an abundance of correlation-driven states at fractional filling as well as robust superconductivity. 
The absence of linear-in-$T$ resistivity without significant reduction of the superconducting transition temperature, provides insights into the origin of both phenomena.
Most remarkably, the hierarchy between integer and fractional filling is absent, indicating that flavor polarization does not play a governing role. The prominence of fractional filling in the small twist angle regime also points towards a longer-range effective Coulomb interaction. 
Combined, our results shed new light on outstanding questions in the field, while establishing the small twist angle regime as a new paradigm for exploring novel flavors of moir\'e physics.}

 \begin{figure*}
\includegraphics[width=0.75\linewidth]{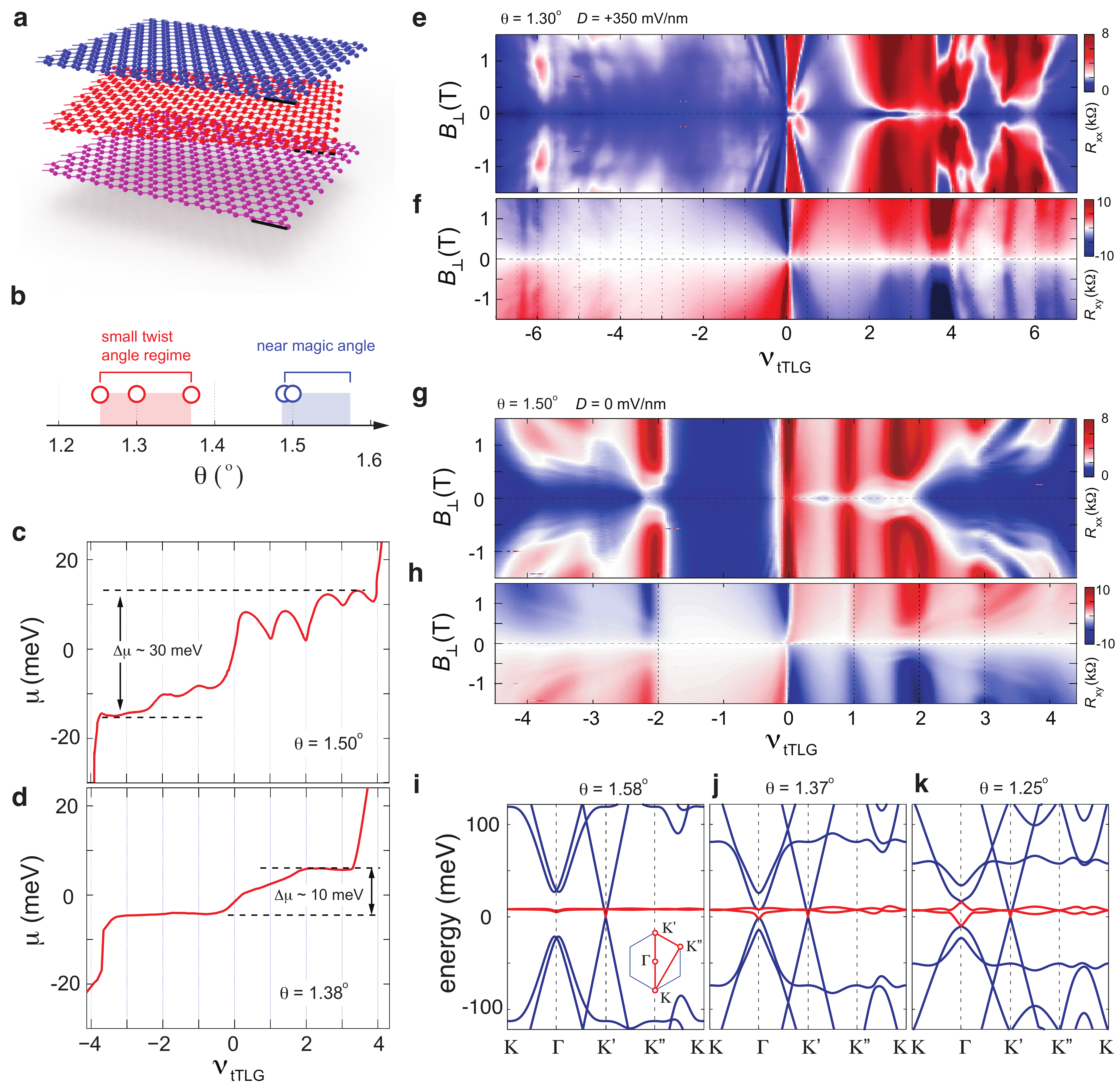}
\caption{\label{fig1}{\bf{Novel flatband condition at small twist angle.}} (a) Schematic of mirror symmetric twisted trilayer graphene. (b) Twist angles where a flat energy band has been observed in tTLG. (c-d) Chemical potential $\mu$ measured at twist angle of (c) $\theta = 1.50^{\circ}$ (sample D) and (d) $1.38^{\circ}$ (sample E), as a function of moir\'e filling $\nu_{tTLG}$. \muu\ jumps abruptly at $\nu_{tTLG} = \pm4$, which corresponds to the transition between the primary and remote moir\'e bands.  (e) \Rxx\ and (f) \Rxy\ as a function of moir\'e filling $\nu_{tTLG}$ and magnetic field $B$ measured at $\theta=1.30^{\circ}$ (sample B).  (g) \Rxx\ and (h) \Rxy\ as a function of moir\'e filling $\nu_{tTLG}$ and magnetic field $B$ measured at $\theta=1.50^{\circ}$ (sample D). (i-k) Calculated band structure of tTLG at different twist angle, see SI 2 \cite{SI} for details.
}
\end{figure*}

\begin{figure*}
{\includegraphics[width=0.95\linewidth]{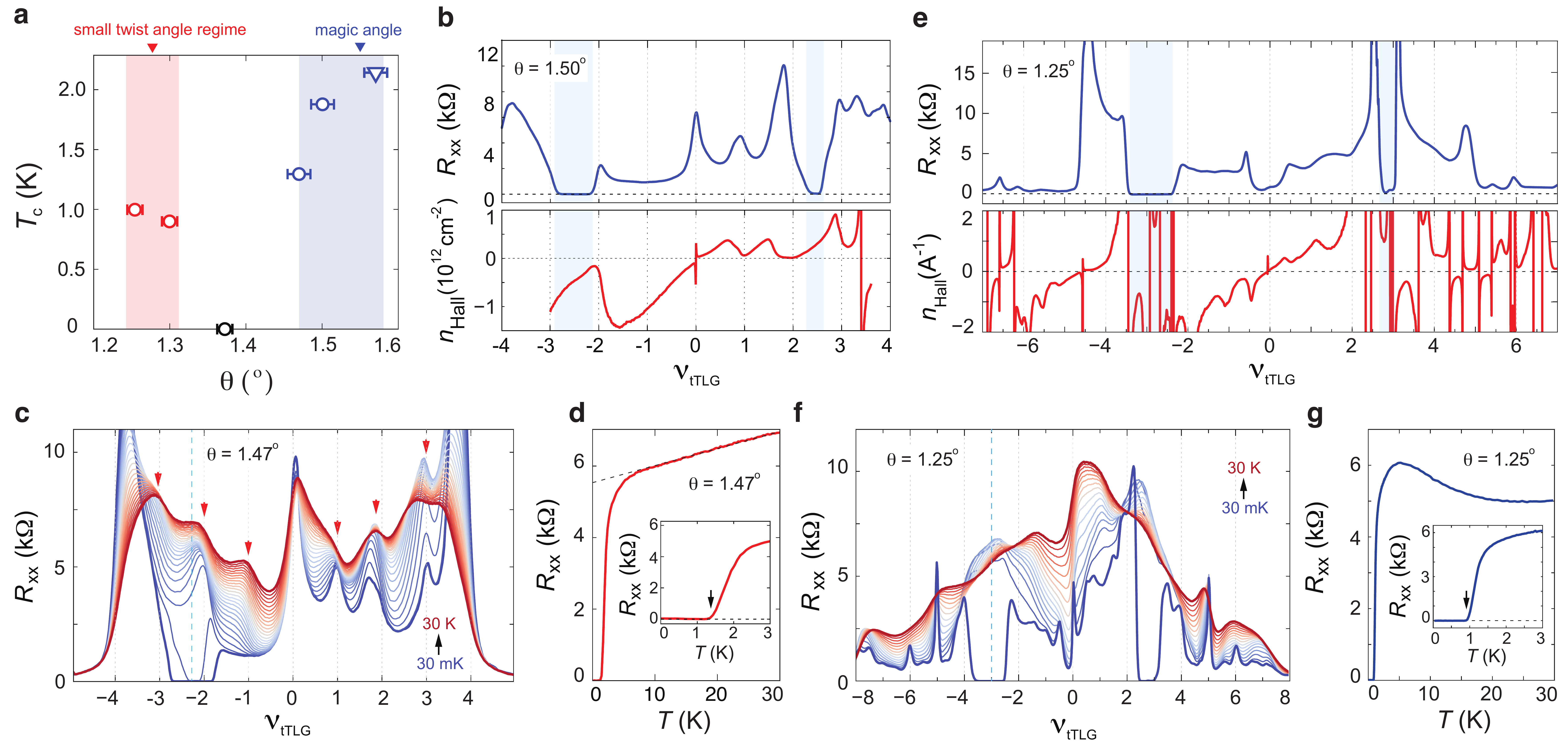}
\caption{\label{figSC} {\bf{Superconductivity.}} (a) Superconducting transition temperature $T_c$  as a function of $\theta$. The triangular data point is taken from Ref.~\cite{Park2021tTLG}. (b) Longitudinal resistance \Rxx\ (top panel) and Hall density $n_H$ (bottom panel) as a function of moir\'e filling $\nu_{tTLG}$ measured from sample D with twist angle $\theta = 1.50^{\circ}$. (c) \Rxx\ as a function of $\nu_{tTLG}$ measured at different temperature from sample C with twist angle $\theta = 1.47^{\circ}$. (d) \Rxx\ as a function of temperature measured at $\nu_{tTLG}=-2.2$, marked by the blue dashed line in panel (c). Inset, \Rxx\ as a function of $T$ around the superconducting transition temperature of $T_c$. (e) Longitudinal resistance \Rxx\ (top panel) and Hall density $n_H$ (bottom panel) as a function of moir\'e filling $\nu_{tTLG}$ measured from sample A with twist angle $\theta = 1.25^{\circ}$. (f) \Rxx\ as a function of $\nu_{tTLG}$ measured at different temperature from sample A ($\theta = 1.50^{\circ}$). (g) \Rxx\ as a function of temperature measured at $\nu_{tTLG}=-3$, marked by the blue dashed line in panel (f). Inset, \Rxx\ as a function of $T$ around the superconducting transition temperature of $T_c$.  }}
\end{figure*}

The interplay of the narrow bandwidth near the magic angle of twisted graphene systems and the Coulomb interaction can give rise to the spontaneous polarization of the internal flavor degrees of freedom of the moir\'e unit cell, given by spin, valley, and the two almost degenerate conduction and valence flat bands. Different forms of this flavor polarization are believed to explain many emergent correlated phenomena in these systems, such as correlated insulators ~\cite{Lu2019SC,Cao2018a,Yankowitz2019SC}, orbital ferromagnets~\cite{Sharpe2019,Serlin2019,Chen20201N2,Polshyn20201N2,Lin2021SOC}, cascade of band resets~\cite{Park2021flavour,Zondiner2020cascade,Wong2020cascade,Kang2021cascades}, resets in the Hall density, and saw tooth patterns in electronic compressibility~\cite{Park2021flavour,Rozen2020pomeranchuk,Saito2021pomeranchuk,Liu2021DtTLG}. 
The dominance of flavor polarization is also consistent with the hierarchical behavior at the magic angle: the most prominent correlation-driven phases appear at integer fillings~\cite{Cao2018a,Cao2018b,Liu2020DBLG,Cao2020DBLG,Chen20201N2,Polshyn20201N2,Park2021tTLG,Hao2021tTLG}, whereas quantum phases at incommensurate fillings tend to be weaker~\cite{Xie2021tblg,Pierce2021tblg}. 
This hierarchy suggests that the Coulomb interaction leaves the translational symmetry of the moir\'e superlattice unbroken, which can be understood, at least phenomenologically, by electronic correlations on the length scale of one moir\'e wavelength.
Importantly, the hierarchy likely also crucially affects the superconducting state as can be seen, e.g., in twisted trilayer graphene (tTLG) where the superconducting region in the phase diagram tracks features in the normal-state Hall density \cite{Park2021tTLG,Hao2021tTLG}.  

In this work, we investigate mirror-symmetric tTLG (Fig.~\ref{fig1}a) in the small twist angle regime and uncover very flat bands in multiple samples in the twist angle range of $1.25^{\circ}$ to $1.38^{\circ}$ (Fig.~\ref{fig1}b, see Table~\ref{DiodeEffect2} for the list of samples studied), with bandwidth even smaller than that of our samples close to the magic angle. For simplicity, we will refer to this range of twist angle as the small twist angle regime (Fig.~\ref{fig1}b).
In this regime, we report a series of robust fractional filling correlated states, providing  evidence for a longer-ranged Coulomb interaction, with electronic correlations occurring on the length scale of multiple moir\'e wavelengths; notably the aforementioned hierarchy between integer and fractional filling is absent.
We will examine the role of longer-ranged Coulomb interaction and of the absence of hierarchy on the phenomenology of tTLG by studying transport responses of correlation-driven phases and superconductivity.
We will focus our discussion here on the twist angle dependence and transport behaviors which are mostly insensitive to the presence of the atomic interface between tTLG and a thin crystal of tungsten diselenide (\WSe). The influence of the tTLG/\WSe\ interface, on the other hand, will be discussed elsewhere.

The remarkably small bandwidth of the primary moir\'e bands in the small twist angle regime away from the magic angle is revealed using chemical potential measurements. 
Fig.~\ref{fig1}c and d plot the evolution of \muu\ across the primary moir\'e band at $\theta=1.50^{\circ}$ and $1.38^{\circ}$, respectively.
The net increase of \muu\ between moir\'e filling of $\nu=\pm4$ offers an experimental definition for the moir\'e bandwidth. Interestingly, the moir\'e bandwidth is less than $10$ meV in the small twist angle regime, which is much smaller compared to the value of $30$ meV near the magic angle. Despite the flat energy band in the small twist angle regime, however, \muu\ appears mostly featureless across the moir\'e energy band. This is in stark contrast compared to the saw tooth pattern in \muu\ observed at $1.50^{\circ}$ (Fig.~\ref{fig1}c). Since this saw tooth pattern is directly linked to spontaneous flavor polarization~\cite{Park2021flavour,Zondiner2020cascade,Wong2020cascade,Kang2021cascades}, the absence of such pattern suggests that the dominant correlated features of the system are not captured by the common picture of flavor polarization at integer fillings.

Magnetotransport data offer more insights. 
The magnetotransport response in the small twist angle regime, as shown in Fig.~\ref{fig1}e-f, exhibits a series of features in both the primary and remote bands with density modulation of fractional moir\'e filling. This is different from the magic angle regime where correlation-driven phases appear at every integer filling.
The $B$-dependence of magnetotransport response allows us to make an interesting observation regarding the underlying band structure. 
Near the magic angle, magnetotransport exhibits a sharply defined transition in $B$ (Fig.~\ref{fig1}g-h), which arises from the coexistence between a flat moir\'e band and a highly dispersive Dirac band.  At small $B$, transport behavior is dominated by the dispersive Dirac band, which gives rise to a series of extra Landau levels emanating from the charge neutrality point (blue  features in Fig.~\ref{fig1}g which are almost horizontal). These Dirac band Landau levels disappear when the Landau gap of the Dirac band exceeds the bandwidth of the flat band at roughly $B=0.5$ T. Notably, this transition offers another measure of the moir\'e bandwidth, $v_F\sqrt{2e\hbar B}\approx 26\,\textrm{meV}$, which is nicely consistent with the measured bandwidth in Fig.~\ref{fig1}c (see Method section for more detailed discussion). On the high-field side of the transition, transport response is dominated by correlation-driven phases in the flatband, which are demonstrated by a series of vertical features that emerge at integer fillings ~\cite{Park2021flavour}. This $B$-induced transition is missing in the small twist angle regime, as shown in Fig.~\ref{fig1}e-f, suggesting that the band structure is modified compared to that of the magic angle. The resulting band structure in the small twist angle regime cannot be viewed as a combination of flat moir\'e band and a dispersive Dirac band.

A continuum model description, which is the trilayer analog of Ref.~\cite{Bistritzer2011}, offers a framework to understand the small bandwidth and magnetotransport response in the small twist angle regime.
In this model, the key parameters determining the twist-angle $\theta$ dependence of the band structure are the intra- ($w_0$) and inter-sublattice ($w_1$) tunneling strengths. For $w_0=w_1=w$, one finds a sharp minimum of the bandwidth as a function of $\theta$. The band structure for the frequently used value $w=110\,\textrm{meV}$~\cite{Bistritzer2011} is shown in Fig.~\ref{fig1}i, which exhibits a minimum bandwidth near the magic angle. This is consistent with previous observations where strong Coulomb correlation near the magic angle leads to a cascade of flavor symmetry breaking at integer filling ~\cite{Park2021tTLG,Hao2021tTLG,Liu2021DtTLG}.  
With these parameters, however, the bandwidth in the small twist angle regime would be much larger than that in Fig.~\ref{fig1}d. One way to reconcile this is to assume that $w=w_0=w_1$ decreases significantly with the twist angle since the magic angle is approximately proportional to $w$.
Another, more natural, explanation is based on using a reduced $w_0 < w_1=w$, which is known \cite{PhysRevResearch.1.013001} to describe relaxation effects. While $w_0/w_1$ in reality also depends on the twist angle, already a fixed $w_0/w_1$ gives rise to minima in the bandwidth at the magic angle and in the small twist angle regime (see Fig.~\ref{figmodel}a in \cite{SI}). 
The band structures calculated using a reduced ratio of $w_0/w_1=0.875$ at $\theta=1.37^{\circ}$ and $1.25^{\circ}$ are shown in Fig.~\ref{fig1}j and k, respectively. 
According to the model, the primary moir\'e bands, highlighted in red, remains relatively flat in the small twist angle regime. Notably,  Fig.~\ref{fig1}j and k exhibit an extra band crossing between the remote and primary bands at the $\Gamma$ point, which is less dispersive compared to the Dirac band at the $K'$ point of the Brillouin zone and becomes increasingly flat with decreasing twist angle.
The presence of the less dispersive crossing at the $\Gamma$ point can explain the suppression of the Landau level features associated with the Dirac band in Fig.~\ref{fig1}e-f.
At the same time, the model indicates that some of the remote bands become less dispersive with decreasing twist angle, making it possible to host correlation-driven phases. Moreover, the energy separation between the primary and the increasingly flat remote bands is shown to be reduced at small twist angle. The energetic proximity of a band with high density of states is likely going to alter the form and impact of the Coulomb interactions in the system, as we will see below.

Having discussed the main phenomenology associated with the energy band structure in the small twist angle regime, we are now in position to discuss their influence on superconductivity. Despite the absence of fermi surface reconstruction,  superconductivity is stable in the small twist angle regime, with transition temperature similar to previous observations near the magic angle (Fig.~\ref{figSC}a) ~\cite{Park2021tTLG,Hao2021tTLG,Liu2021DtTLG}.
We will show that transport behaviors associated with the superconducting phase in the small twist angle regime are distinct compared to previous observations near the magic angle, which highlights the possibility of a new superconducting phase.
Published experimental literature studying magic-angle graphene moir\'e systems linked the superconducting phase to several transport responses arising from fermi surface reconstruction at low temperature and fluctuating flavor moments at high temperature~\cite{Park2021flavour,Rozen2020pomeranchuk,Saito2021pomeranchuk,Liu2021DtTLG}. 
This can also be seen in our data in Fig.~\ref{figSC}b for $\theta$ close to the magic angle: the superconducting phase is associated with a fermi surface reconstructed by flavor polarization, which is evidenced  by the apparent reset in the Hall density at $|\nu_{tTLG}|=2$.
The tendency to spontaneously polarize some flavor degrees of freedom also gives rise to strong fluctuations in local moments at high temperature. This is manifested in a series of resistance peaks at every integer filling  (marked by red vertical arrows in Fig.~\ref{figSC}c). Local moments persist to around $T = 100$ K, roughly the same scale as the Coulomb interaction near the magic angle ~\cite{Saito2021pomeranchuk}. 
In the small twist angle regime, however, the influence of flavor polarization is substantially suppressed. At low temperature, the superconducting phase is not accompanied by Hall density resets (Fig.~\ref{figSC}e). Instead, a series of vHSs with diverging density of states is observed in the density range of the superconducting phase. This is in stark contrast with previous observations near the magic angle, where superconductivity is shown to diminish near van Hove singularities (vHSs)  ~\cite{Park2021tTLG}.  
At high temperature, the longitudinal resistance \Rxx\ evolves monotonically as the sample is tuned away from the charge neutrality point. The lack of density modulation with a periodicity of one moir\'e filling, as shown in Fig.~\ref{figSC}f, provides further confirmation that the local flavor moment is suppressed in the small twist angle regime.

It is also worth pointing out that the linear-in-T dependence of $R_{xx}$ is absent in the small twist angle regime over the entire superconducting density range. Fig.~\ref{figSC}g shows that longitudinal resistance decreases monotonically with increasing temperature, which is in stark contrast with 
the linear-in-T behavior that has been universally linked with superconductivity in magic-angle graphene moir\'e systems (Fig.~\ref{figSC}d and Fig.~\ref{figRT}) ~\cite{Cao2020strange,Jaoui2021quantum,Polshyn2019phonon}. Our finding, therefore, suggests that the origin of superconductivity in the small twist angle regime is decoupled from the mechanism underlying the linear-in-T temperature dependence of \Rxx.
At the same time, the comparison between different twist angle regimes offers new insights into the linear-in-T behavior. 
We first note that the transition temperature of the superconducting phase is mostly insensitive to twist angle, despite significantly altered band structure (inset of Fig.~\ref{figSC}d and g).
If we assume that superconductivity comes from electron-phonon pairing or is at least crucially stabilized by it ~\cite{Liu2021DtBLG,Liu2021DtTLG}, the approximately identical $T_c$ indicates that the electron-phonon coupling is roughly the same. As such, the absence of the  linear-in-$T$ resistance in the small twist angle regime suggests that the said phenomenon does not come from electron-phonon coupling alone~\cite{Polshyn2019phonon}. 
Together with our observation of suppressed flavor polarization in the small twist angle regime, a natural interpretation is that the linear-in-T behavior arises from the fluctuation of flavor moments at high temperature~\cite{Saito2021pomeranchuk}.

\begin{figure*}
\includegraphics[width=0.82\linewidth]{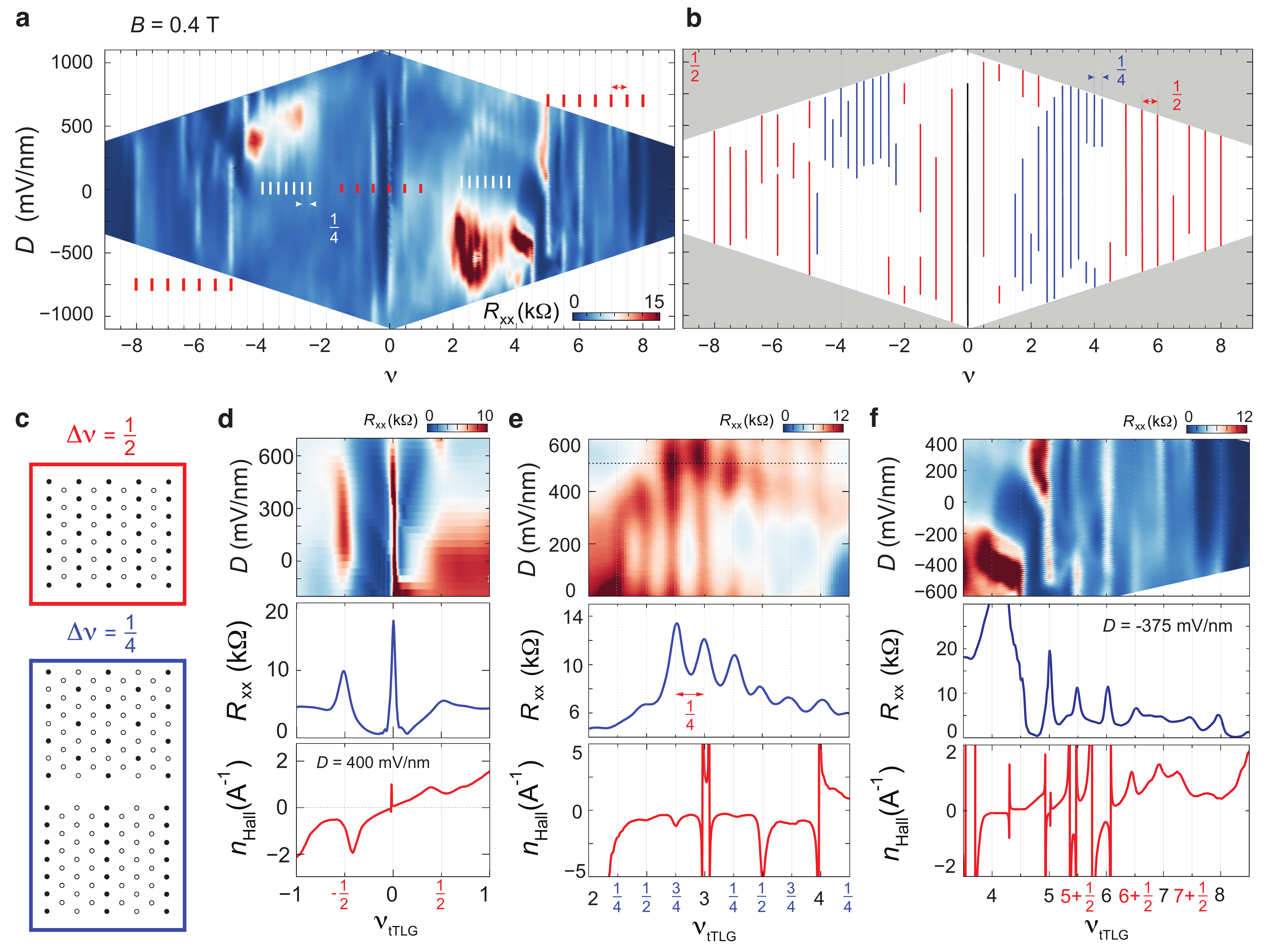}
\caption{\label{figDW} {\bf{Density wave instabilities.}}  (a) Longitudinal resistance \Rxx\ as a function of $D$ and moir\'e filling $\nu_{tTLG}$ measured at $B=0.4$ T and $T= 20$ mK. (b) Schematic for the $\nu_{tTLG}$-$D$ phase diagram where the location for resistive peaks are marked by solid red and blue lines. Both blue and red lines track constant moir\'e filling $\nu_{tTLG}$. Blue (red) lines emerge with a periodicity of $1/4$ ($1/2$) in moir\'e filling. (c) Schematics of the spatial charge distribution of DW states with periodicity of $1/2$ (top panel) and $1/2$ (bottom two panels) in moir\'e filling. (d-f) $\nu_{tTLG}$-$D$ map of \Rxx\ (top panel), \Rxx\ (middle panel) and Hall density $n_{\text{Hall}}$ (bottom panel) as a function of $\nu_{tTLG}$ in the moir\'e filling range of (d) $-1 < \nu_{tTLG} < 1$, (e) $2 < \nu_{tTLG} < 4+\frac{1}{4}$, and (f)  $3.5 < \nu_{tTLG} < 8.5$.    All measurements are performed in sample A.
}
\end{figure*}

\begin{figure*}
\includegraphics[width=0.76\linewidth]{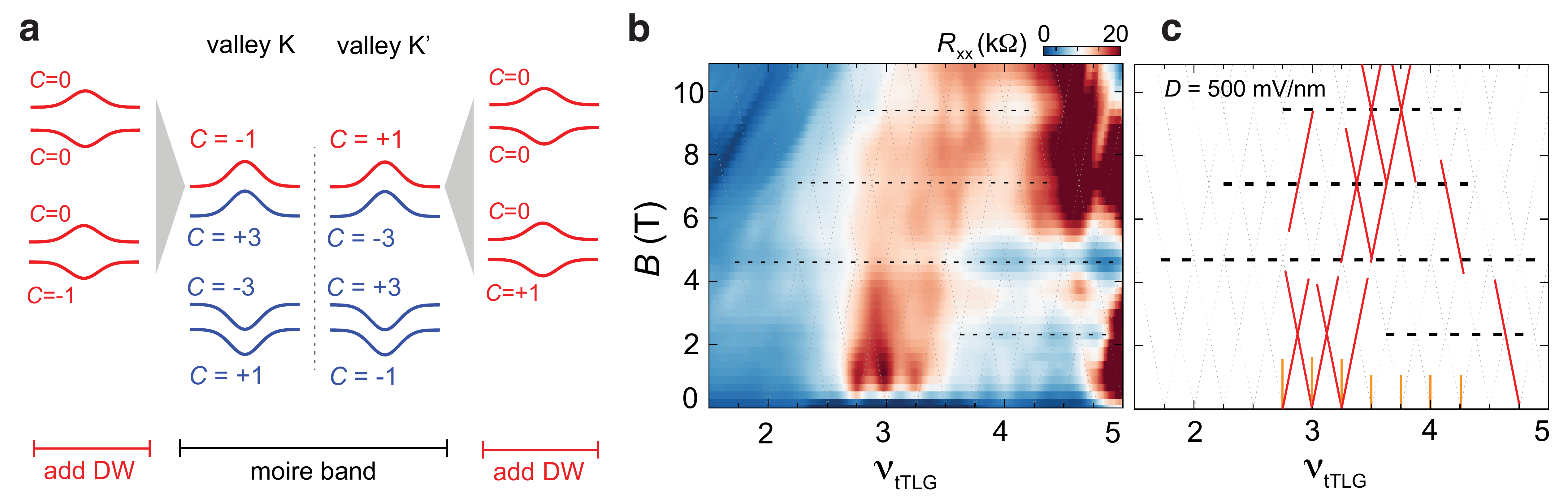}
\caption{\label{figCI} {\bf{Incompressible states with a perpendicular magnetic field.}}  (a) Possible evolution of the spin-orbit-coupled moir\'e bands in the tBLG sector of tTLG with 1/4 DW order. (b) $\nu_{tTLG}$-$B$ map of \Rxx\ around half-filled moir\'e band measured at $D = 500$ mV/nm. (c) Schematic summarizing the most prominent features in panel (b).  Red solid lines highlights broken symmetry Chern insulators (SBCI) with $s$ taking quarter-integer values.  Yellow vertical lines correspond to DW states with $t=0$. Horizontal dashed lines mark every quarter magnetic flux per moir\'e cell. 
All measurements are performed in sample A.
}
\end{figure*}

The suppression of flavor polarization offers strong indication that the form of Coulomb interaction has been altered significantly in the small twist angle regime compared to the magic angle. 
In the following we will investigate the nature of Coulomb interaction  by studying the behavior of correlation-driven phases in the small twist angle regime. We will suppress the influence of the superconducting phase using a small external magnetic field and focus on 
sample A (see Table~\ref{DiodeEffect2}). 
In the $\nu_{tTLG}$-$D$ plane of sample A at $\theta = 1.25^{\circ}$, longitudinal resistance \Rxx\ exhibits a series of peaks at partial band filling (Fig.~\ref{figDW}a). 
The positions of the most pronounced peaks are labeled in the schematic $\nu_{tTLG}$-$D$ phase diagram in Fig.~\ref{figDW}b. 
Without using any fitting parameters, all resistance peaks in the $\nu_{tTLG}$-$D$ phase diagram, across both the flat and remote bands, are shown to match integer multiples of $1/2$ or $1/4$ moir\'e fillings.   The density modulation of fractional moir\'e filling points towards spontaneously breaking the moir\'e translation symmetry and the formation of density wave (DW) instabilities~\cite{Xu2020DW,Xie2021tblg}, which are driven by long-range Coulomb correlations. 
For simplicity, we will refer to these DW states according to the periodicity of their emergence as a function of $\nu_{tTLG}$, e.g., the blue and red solid lines in Fig.~\ref{figDW}b represent $1/4$ and $1/2$ DW states, respectively.

The broken moir\'e translation symmetry is also reflected by the Hofstadter butterfly spectrum, where high symmetry points in the recursive structure of the Hofstadter butterfly spectrum is demonstrated by horizontal features of maximum in longitudinal conductance and sign-reversals in Hall conductance (Fig.~\ref{f:spectrum} and Fig.~\ref{f:sampleA2}) ~\cite{Dean2013hofstadter,Hunt2013hofstadter,Ponomarenko2013hofstadter}. The most pronounced horizontal feature occurs around $B = 5$~T, which corresponds to half a magnetic flux every four moir\'e unit cells, $4\phi/\phi_{0}=0.5$. Here $\phi$ is the magnetic flux per moir\'e unit cell, and $\phi_0$ is the magnetic flux quantum. This is in excellent agreement with the periodicity of the correlated states, which show density modulation of $1/2$, $1/4$ and $1/8$ moir\'e filling (see Fig.~\ref{figDW}c, Fig.~\ref{figSBCI}, and Fig.~\ref{f:linecut}) ~\cite{SI}. In addition, the presence of these high symmetry points throughout the entire density range in the $B$-$\nu_{tLG}$ plane (Fig.~\ref{figDW}) offers strong evidence that the moir\'e wavelength and twist angle are well-defined throughout the uniform sample.

Theoretically, the emergence of 1/4 (1/2) DW order, with  spatial modulation illustrated on the triangular lattice in Fig.~\ref{figDW}c, 
should lead to a complex reconstruction of the moir\'e bands into 4 (2) sub bands each, splitting the van Hove singularity (vHs) of the unreconstructed band into multiple peaks (see SI 3 and Fig.~\ref{f:theoryplots}d ~\cite{SI} for a demonstration). Fig.~\ref{figCI}a illustrates one of the possible scenarios for splitting moir\'e bands into DW sub bands, where one of the sub bands inherits the Chern number of the original spin-orbit-split \cite{Lin2021SOC} bands. The formation of DW sub bands is expected to induce complex signatures in the $\nu_{tTLG}$-dependence of the Hall density $n_{Hall}$, which we indeed observe: 
at $\nu_{tTLG}=\pm 1/2$, resistance peaks are accompanied by resets 
in Hall density, where $n_{Hall}$ diminishes at the onset of DW order (Fig.~\ref{figDW}d). The fact that the density of mobile charge carriers is suddenly reduced is indicative of, at least, a partial gap opening in the Brillouin zone due to DW order. 
At higher densities, we also observe multiple vHs-like features in $n_{Hall}$ with diverging density of states (Fig.~\ref{figDW}e-f).
A more direct demonstration of the DW-induced band reconstruction is provided by the behavior of incompressible states in the presence of a perpendicular magnetic field. Linearly-dispersing incompressible states in the quantum Hall effect regime can be classified using a pair of quantum numbers ($t, s$) from the Diophantine equation $\nu = t \phi/\phi_{0} + s$, where $\nu$ is the filling factor $\nu_{tTLG}$ at the incompressible state ~\cite{Xie2021tblg,Spanton2018}. According to the Streda formula, the slope $t$ is equal to the state's Chern number, $C=t$~\cite{Streda_1982}, and $s$ describes the moir\'e filling of the low-field DW state from which the incompressible state develops. In this format, incompressible states in the density range of $1 < \nu_{tTLG} < 4$ are described by $(t,s) = (2M,N/4)$, where $M$ and $N$ take integer values. While at first sight the distribution of the Chern numbers in Fig.~\ref{figCI}b seems inconsistent with the presence of DW states with $t=0$ at every $s=N/4$, we note that $C=0$ can be ensured by coherently occupying two DW
bands with opposite valley indices, see SI 3 \cite{SI}. The emergence of additional fans at $s=N/4$ with $C=2M\neq 0$, defining symmetry-broken Chern insulators (SBCIs), provides further confirmation that new Fermi surfaces form at every $1/4$ filling.

Most interestingly, 
correlation-driven phases in the small twist angle regime, including DW and SBCI, demonstrate a unique hierarchical behavior.
For instance, in the density range near half-filling of the moir\'e band, correlation-driven phases with fractional and integer value of $s$ show similar transport responses  (Fig.~\ref{figDW}e and Fig.~\ref{figCI}b). 
This is distinct compared to the behavior near the magic angle, where the most prominent phases develop at integer values of $s$ ~\cite{Pierce2021tblg,Xu2020DW,Xie2021tblg}.
The lack of hierarchy provides another strong indication that flavor symmetry breaking is not the dominant phenomenon governing the system's phase diagram.

Combined together, our findings shed light on a new flavor of correlation-driven phases and superconductivity in the small twist angle regime. The key to understand the distinct behaviors in different twist angle regimes lies in the influence of the modified band structure and the origin of longer-range Coulomb correlation.
Near the magic angle, the measured bandwidth value, 
$\Delta$ \muu $\sim 30$ meV, 
is likely enhanced by Coulomb-driven fermi surface reconstruction whereas the unreconstructed band is much narrower. On the other hand, since flavor symmetry breaking and fermi surface reconstruction are absent in the small twist angle regime, the measured bandwidth of
$\Delta$ \muu $\sim 10$~meV  
is likely much closer to the non-interacting value.  As such, the bare moir\'e bands may be more dispersive in the small twist angle regime compared to the magic angle. As the bands become more dispersive, it is natural to expect that the Coulomb interaction experiences less metallic screening and becomes longer-ranged. Following this, we argue that our findings are consistent with an intermediate regime in twist angle/band dispersion whereby correlated phases are stable, yet electronic correlations occur on the length scale of multiple moir\'e wavelengths, stabilizing an abundance of DW and SBCI phases across the primary and remote moir\'e bands. Together with the observation of the zero-field superconducting diode effect \cite{SCPaper,2021arXiv211209115S} in tTLG below the magic angle, this establishes the small twist angle regime as a promising novel playground for exotic many-body physics.

\section{Method}

\textbf{tTLG samples:}  
Five separate samples are studied in this work, which all contain mirror symmetric tTLG. We list their twist angle, geometry and the transition temperature of the superconducting phase in Table~\ref{DiodeEffect2}. In sample A, C and E, a few-layer \WSe\ crystal is stacked  on top of tTLG, whereas tTLG in sample B and D are fully encapculated by hexagonal boron nitride crystals and the heterestructures do not contain \WSe.   
In all samples, tTLG and the hBN substrate are maximally misaligned to minimize the influence of the hexagonal boron nitride (hBN) substrate (see Fig.~\ref{f:fab}) ~\cite{SI}.

\begin{table}[h]
\begin{center}
\caption{We list properties of five twisted trilayer graphene samples investigated in this work. The two samples near the magic angle, sample C and D, exhibit behaviors consistent with published literature, including fermi surface reconstruction at integer moir\'e filling (Hall density reset, resistance peak, emergence of extra Landau fan, saw tooth pattern in chemical potential, linear-in-T behavior at high temperature). The two samples with the smallest twist angle, Sample A and B, demonstrate drastically different behaviors as discussed in the main text. For example, Hall density measurement shows many vHSs across both the primary and remote bands. Resistance peak and extra Landau fan emerge from fractional fillings. In the density range of the superconducting phase, linear-in-T behavior is absent at high temperature. Sample E appears to be an intermediate sample. It has a twist angle of $\theta = 1.38 ^{\circ}$. On the one hand, correlation-driven phases, along with a series of vHSs are observed in the primary and remote bands; on the other hand, the phase space is not dominated by correlation-driven phases at fractional filling. Most interestingly, no superconductivity is observed in sample E. 
}
\label{DiodeEffect2}
\begin{ruledtabular}
 \begin{tabular} {ccccccccc} 
\multicolumn{1}{c}{Sample} & \multicolumn{1}{c}{$\theta$} & \multicolumn{1}{c}{\WSe} & \multicolumn{1}{c}{$T_c$ (K)}  \\  \hline
A & $1.25^{\circ}$ & \cmark &  $ 1$     \\
 \hline 
B & $1.31^{\circ}$ & \xmark &  $0.9$   \\
 \hline
C & $1.47^{\circ}$ & \cmark &  $1.4$     \\
  \hline
D & $1.50^{\circ}$ & \xmark &  $1.9$     \\
    \hline
E & $1.38^{\circ}$ & \cmark &  $0$ (no superconductivity)    \\
 \end{tabular}
\end{ruledtabular}
\end{center}
\end{table}

\textbf{Bandwidth calculated based on magnetotransport data:}
For a Dirac band, the Landau level gap near the CNP is defined as $\Delta_{LL}= v_F\sqrt{2e\hbar B}$. When $\Delta_{LL}$ exceeds the bandwidth of the moir\'e band, there are no free charge carriers in the Dirac band and transport response is dominated by the moir\'e flatband. As a result, the magnetic field value where the influence of the Dirac band disappears provides a characterization for the moir\'e bandwidth. For instance, Fig.~\ref{fig1}g-h show that a $B$-field induced transition at $\nu_{tTLG}=+2$ occurs at $B = 0.5$ T, which corresponds to a bandwidth of $\approx 26\,\textrm{meV}$, in excellent agreement with the chemical potential measurement in Fig.~\ref{fig1}c. 

\textbf{Twist angle mismatch:} we address possible relations to twist angle mismatch between top/middle and middle/bottom graphene layers in our tTLG sample. While a reconstruction of the moir\'e bands could theoretically be induced by a moir\'e pattern of moir\'e unit cells, resulting from a fine-tuned twist-angle mismatch between the top and bottom graphene layers \cite{ColumbiaSTMttlg}, we believe that this scenario provides a much less natural interpretation of our findings than interaction-induced DW instabilities. 

First, in the presence of a small twist angle mismatch between top and bottom graphene layers, transport responses of tTLG samples, such as Hall density and magnetoresistance, can be well explained with a single particle picture, as shown in Fig.~\ref{f:misalign}. In these samples, the small twist angle mismatch gives rise to two weak satellite peaks near the charge neutrality point (marked by vertical red lines in Fig.~\ref{f:misalign}b), which are the only identifiable resistance features within the moir\'e band ($-4 < \nu < +4$). The twist angle mismatch can be identified using these resistance peaks. Notably, superconductivity is always suppressed in the presence of twist angle mismatch in tTLG samples, as shown in Fig.~\ref{f:misalign}e-f. These behaviors are in stark contrast with our observations in the small twist angle regime. 

Second, unlike moir\'e-induced band reconstruction which will give rise to an approximately temperature independent gap, 
the vHSs associated with 1/4 DW states disappear with increasing temperature at $T \sim 1$ K (Supplemental Fig.~\ref{HallT}) ~\cite{SI}. This points towards a Coulomb-driven origin. In a separate work \cite{SCPaper}, we show that superconductivity is stabilized in this sample at $B=0$. If twist angle mismatch gave rise to a 4-fold enlarged moir\'e supercell, superconductivity would be expected to exhibit a density modulation with $1/4$ filling periodicity. However, despite the emergence of superconductivity in regions of the $\nu_{tTLG}$-$D$ phase diagram where DW states are present at small magnetic fields, its robustness is largely insensitive to the density modulation. 

Finally, a density modulation of $1/2$ and $1/4$ filling has been observed in all samples with twist angles spanning a range of $1.25^{\circ}$ to $1.5^{\circ}$ degrees ~\cite{SI}, suggesting that density wave instability is common for tTLG samples. An accidental twist angle mismatch is unlikely to yield such reproducibility.  
This is further supported by the following observations: (i) the location of all correlated insulators throughout the density range is accurately matched with integer multiple of $1/2$ and $1/4$  without any fitting parameter;  (ii) a high symmetry point in the Hofstadter spectrum is observed at $\phi/\phi_{0}=1/4$  over the full density range. These observations highlights a well-defined moir\'e wavelength and excellent homogeneity of the tTLG sample, whereas a twist angle mismatch is known to give rise to  sample inhomogeneity ~\cite{ColumbiaSTMttlg}.

Combined, our findings suggest that the insulating phases at fractional fillings are correlation-driven DW instabilities, which are intrinsic to the tTLG/\WSe\ heterostructure, rather than a result of twist-angle mismatch.

\section*{Acknowledgments}
J.I.A.L and J.L are supported by NSF DMR-2143384. P.S. acknowledges support from the Brown University Undergraduate Teaching and Research Awards. Device fabrication was performed in the Institute for Molecular and Nanoscale Innovation at Brown University. 
H. D. Scammell acknowledges funding from ARC Centre of Excellence FLEET.
K.W. and T.T. acknowledge support from the Elemental Strategy Initiative
conducted by the MEXT, Japan (Grant Number JPMXP0112101001) and  JSPS
KAKENHI (Grant Numbers 19H05790, 20H00354 and 21H05233).

\bibliography{Li_ref}

\newpage

\newpage
\clearpage

\pagebreak

\begin{widetext}
\section{Supplementary Materials}

\begin{center}
\textbf{\large A new flavor of correlation and superconductivity in small twist-angle trilayer graphene}\\
\vspace{10pt}

Phum Siriviboon, Jiang-Xiazi Lin, Harley D. Scammell, Song Liu, Daniel Rhodes, K. Watanabe, T. Taniguchi, James Hone, Mathias S. Scheurer, and J.I.A. Li$^{\dag}$

\vspace{10pt}
$^{\dag}$ Corresponding author. Email: jia$\_$li@brown.edu
\end{center}

\noindent\textbf{This PDF file includes:}

\noindent{Supplementary Text}

\noindent{Materials and Methods}

\noindent{Figs. S1 to S19}

\noindent{References (44-47)}

\renewcommand{\vec}[1]{\boldsymbol{#1}}

\def\theequation{S\arabic{equation}}
\renewcommand{\thefigure}{S\arabic{figure}}
\setcounter{figure}{0}
\setcounter{equation}{0}

\newpage

\subsection{SI 1: twist angle dependence}

\begin{table}[h]
\begin{center}
\caption{We summarize the main observations in different twist angle regimes. The twist angle dependence of the zero-field superconducting diode effect is discussed in Ref. ~\cite{SCPaper}.
}
\label{DiodeEffect2}
\begin{ruledtabular}
 \begin{tabular}{ccc} 
 \\
\multicolumn{1}{c}{} & \multicolumn{1}{c}{Near the magic angle} & \multicolumn{1}{c}{small twist angle regime}   \\  
\\
\hline
\\
Superconductivity & \cmark & \cmark    \\
\\
 \hline 
\\
Flavor polarization & \cmark & \xmark   \\
\\
 \hline
\\
Linear-in-T behavior & \cmark & \xmark      \\
\\
  \hline
\\
Zero-field superconducting diode effect & \xmark & \cmark      \\
\\
    
 \end{tabular}
\end{ruledtabular}
\end{center}
\end{table}

\begin{figure*}[h]
{\includegraphics[width=0.8\textwidth,clip]{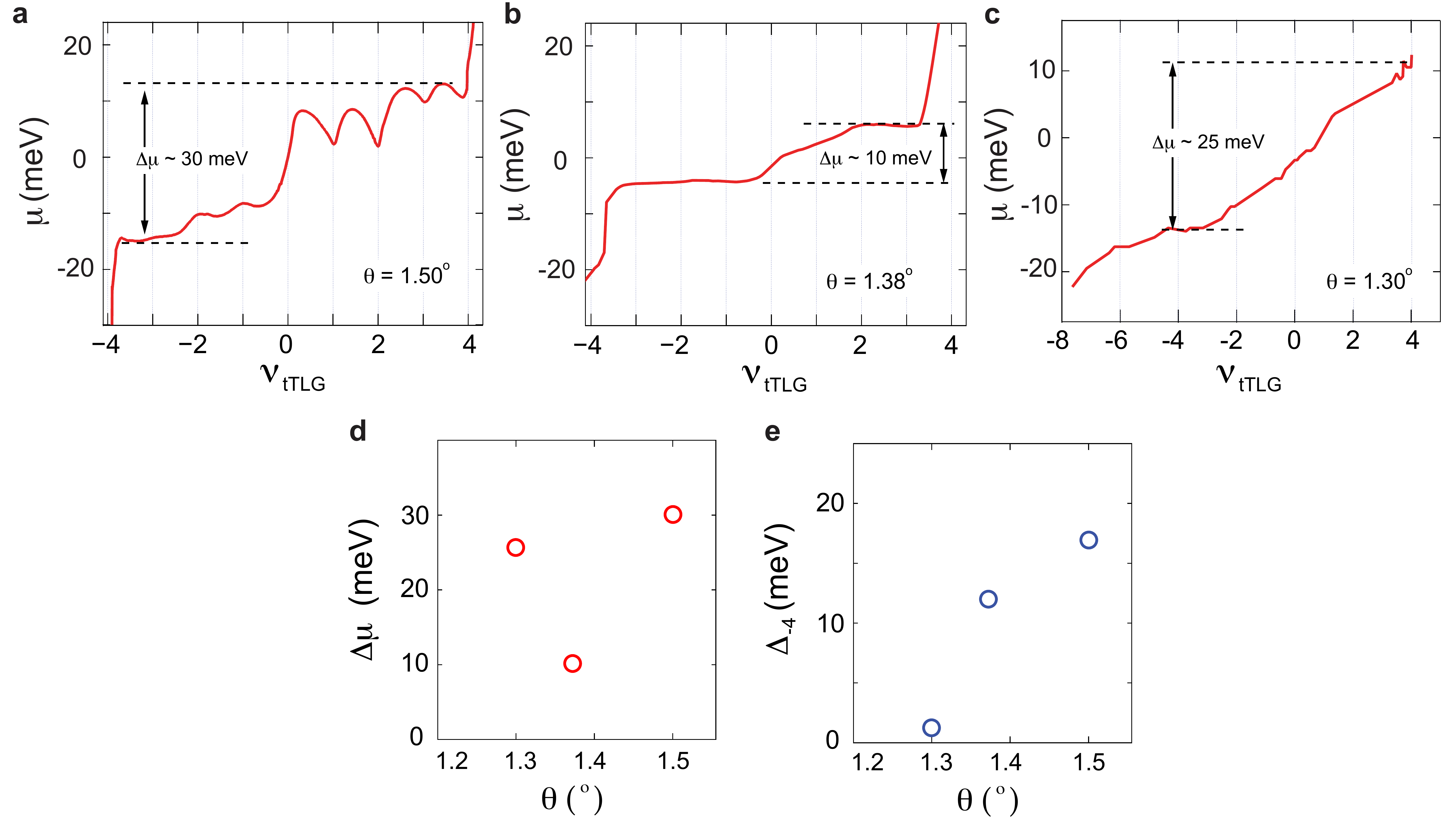}
\caption{\label{figmu}\textbf{Band structure vs $\theta$ } (a-c) Chemical potential $\mu$ versus $\nu_{tTLG}$ across the moir\'e bands measured at different $\theta$. (d) Moir\'e bandwidth, defined by the increase in \muu\ across the primary moir\'e band $\Delta \mu$, as a function of twist angle $\theta$. The measured $\Delta \mu$ at $\theta=1.50^{\circ}$ is likely enhanced by Coulomb correlation and the unreconstructed bandwidth is likely much smaller. However, the bandwidth at $\theta=1.50^{\circ}$ is smaller compared to the model calculation using expected values for $w_0$ and $w_1$. This very flat band can be explained using modified values of $w_0$ and $w_1$. (e) The jump in chemical potential at $\nu = -4$, which characterizes the energy gap between the primary and remote energy bands. At $1.5^{\circ}$ and $1.38^{\circ}$, \muu\ exhibits a sharp jump at $\nu=\pm4$. This is because the dispersive Dirac band has a much smaller density of state compared to the flat moir\'e band. At $1.30^{\circ}$, the jump in \muu\ disappears, which is consistent with the calculated band structure showing a flat band crossing at the $\Gamma$ point (Fig.~\ref{fig1}k).  }}
\end{figure*}

\begin{figure*}[h]
{\includegraphics[width=0.4\textwidth,clip]{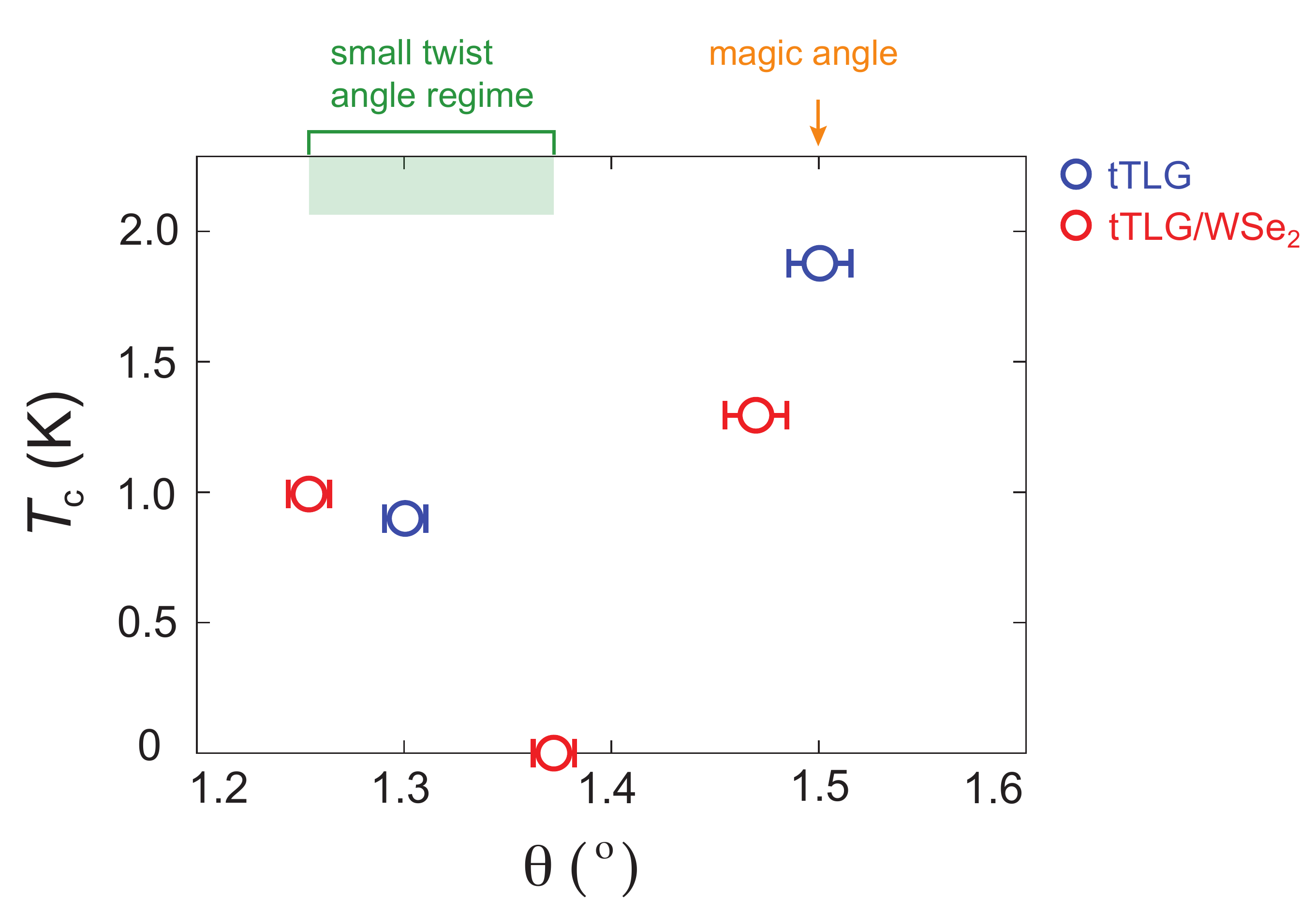}
\caption{\label{figTc}\textbf{Superconducting transition temperature $T_c$ } as a function of $\theta$ measured from five tTLG samples investigated in this work.  The absence of superconductivity in sample E  appears to indicate a transition in the ground state order, which would be indicative of distinct superconducting phases between different twist angle regimes. However, it is worth noting that the behavior of twisted graphene samples are often influenced by sample details. We cannot rule out the possibility of sample E being an outlier without further experimental confirmation. }}
\end{figure*}

\begin{figure*}[h]
{\includegraphics[width=1\textwidth,clip]{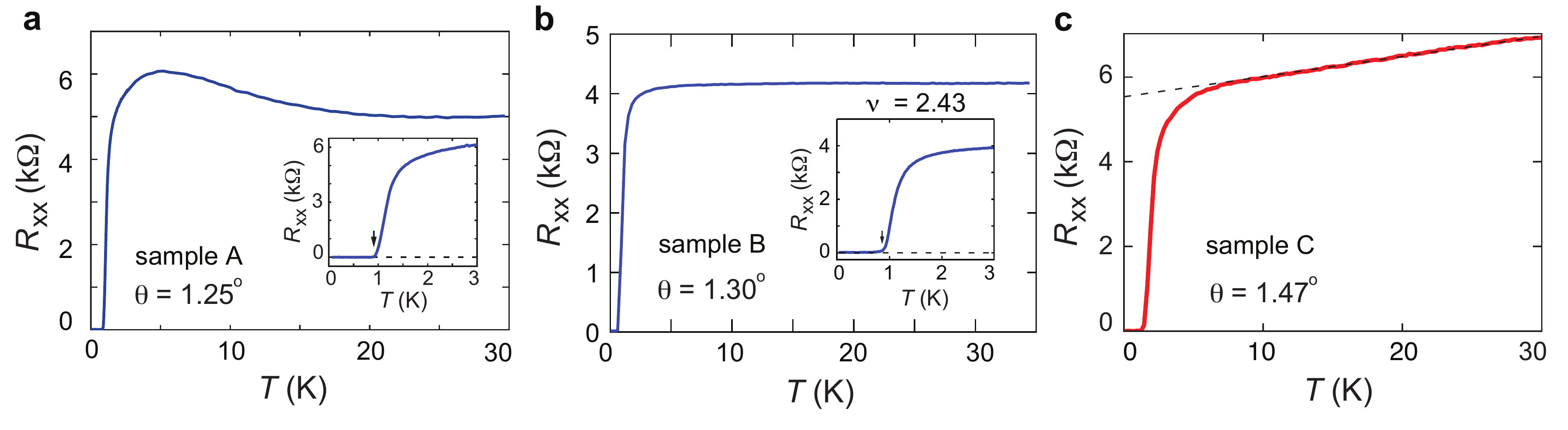}
\caption{\label{figRT}\textbf{Twist angle dependence of linear-in-T behavior} Longitudinal resistance \Rxx\ as a function of temperature measured from (a) sample A at $\theta = 1.25^{\circ}$, (b) sample B at $\theta = 1.30 ^{\circ}$ and (c) sample C at $\theta = 1.50 ^{\circ}$. The measurement is performed at the optimal doping of the superconducting phase. In sample A and B, \Rxx\ does not show clear linear-in-T behavior over any temperature range. All three samples show  sharp superconducting transitions with $T_c \approx 1$ K for sample A and B, and $T_c \approx 1.5$ K for sample C.   }}
\end{figure*}

\newpage

\subsection{SI 2: Band structure calculations}

\noindent \textbf{Noninteracting model.} Here we will state the detailed form of the non-interacting Hamiltonian we use for twisted trilayer graphene in proximity to a WSe$_2$ layer. 
We describe the non-interacting bands of the trilayer graphene system within a continuum model; using an extension of the Bistritzer-MacDonald (BM) model \cite{Bistritzer2011} which accounts for the three layers. 

To define the Hamiltonian, let $c_{\vec{k};\rho,l,\eta,s,\vec{G}}$ denote the operator annihilating an electron at crystalline momentum $\vec{k}$ in the moir\'e Brillouin zone (MBZ), of spin $s=\uparrow,\downarrow$, in sublattice $\rho=A,B$ and valley $\eta = \pm$ of the microscopic graphene sheets, within layer $l=1,2,3$, and with reciprocal lattice (RL) vector $\vec{G} = \sum_{j=1,2} n_j \vec{G}_j$, $n_j \in \mathbbm{Z}$ of the moir\'e lattice. We will use the same symbol with subscript $j=0,1,2,3$ for Pauli matrices and the associated quantum numbers. 

It is convenient \cite{Khalaf2019} to perform a unitary transformation in layer space,
\begin{equation}
    c_{\vec{k};\rho,l,\eta,s,\vec{G}} = V_{l,\ell} \psi_{\vec{k};\rho,\ell,\eta,s,\vec{G}}, \qquad V=\frac{1}{\sqrt{2}} \begin{pmatrix} 1 & 0 & -1 \\ 0 & \sqrt{2} & 0 \\ 1 & 0 & 1 \end{pmatrix} \label{TrafoToMirrorEigenbasis}
\end{equation}
that decomposes the system into mirror-even, $\ell = 1,2$, and mirror-odd, $\ell = 3$, subspaces. Without spin-orbit coupling (SOC) and displacement field, these subspaces will be decoupled as follows from mirror symmetry and can be seen explicitly below. After this transformation, the continuum model is
\begin{equation}
    H_0^{\text{Full}} = \sum_{\vec{k} \in \text{MBZ}} \sum_{\rho,\rho'=A,B} \sum_{\ell,\ell'=1,2,3} \sum_{\eta=\pm} \sum_{s=\uparrow,\downarrow}  \sum_{\vec{G},\vec{G}' \in \text{RL}} \psi^\dagger_{\vec{k};\rho,\ell,\eta,s,\vec{G}} \left(h_{\vec{k},\eta}\right)_{\rho,\ell,\vec{G};\rho',\ell',\vec{G}'} \psi^\pdagger_{\vec{k};\rho',\ell',\eta,s,\vec{G}'}, \label{FullContinuumModel}
\end{equation}
where 
\begin{equation}
    h_{\vec{k},\eta} = h^{(g)}_{\vec{k},\eta} + h^{(t)}_{\vec{k},\eta} + h^{(D)}_{\vec{k}} + h^{(\text{SOC})}_{\vec{k},\eta}. \label{TheBandHamiltonian}
\end{equation}
Here the contributions are: graphene kinetic terms $h^{(g)}_{\vec{k},\eta}$, interlayer tunnelling $h^{(t)}_{\vec{k},\eta}$, displacement field $h^{(D)}_{\vec{k}}$, and proximity-induced SOC $h^{(\text{SOC})}_{\vec{k},\eta}$ due to the WSe$_2$ layer. 

In the mirror basis, the decoupled graphene kinetic terms are
\begin{align}
    \left(h^{(g)}_{\vec{k},+}\right)_{\rho,\ell,\vec{G};\rho',\ell',\vec{G}'} &= \delta_{\ell,\ell'} \delta_{\vec{G},\vec{G}'} v_F (\vec{\rho}_{\theta_\ell})_{\rho,\rho'} \left(\vec{k} + \vec{G} - (-1)^\ell \vec{q}_{1}/2 \right), \label{DiracCones} \\ \left(h^{(g)}_{\vec{k},-}\right)_{\rho,\ell,\vec{G};\rho',\ell',\vec{G}'} &= \left(h^{(g)}_{-\vec{k},+}\right)^*_{\rho,\ell,-\vec{G};\rho',\ell',-\vec{G}'},
\end{align}
where $\vec{\rho}_{\theta} = e^{i \theta \rho_3/2} \vec{\rho} e^{-i \theta \rho_3/2}$, and $\vec{q}_1$ connecting the K and K' points in the MBZ. Moreover, in this basis, the Hamiltonian
\begin{align}
    \left(h^{(t)}_{\vec{k},+}\right)_{\rho,\ell,\vec{G};\rho',\ell',\vec{G}'} &= \sqrt{2} \begin{pmatrix} 0 & (T_{\vec{G}-\vec{G}'})_{\rho,\rho'} & 0 \\  (T_{\vec{G}'-\vec{G}}^*)_{\rho',\rho} & 0 & 0 \\ 0 & 0 & 0 \end{pmatrix}_{\ell,\ell'}, \\  \left(h^{(t)}_{\vec{k},-}\right)_{\rho,\ell,\vec{G};\rho',\ell',\vec{G}'} &= \left(h^{(t)}_{-\vec{k},+}\right)^*_{\rho,\ell,-\vec{G};\rho',\ell',-\vec{G}'},
\end{align}
which accounts for the tunnelling modulated on the moir\'e lattice, only couples the mirror-odd sectors, as required by symmetry. Here we use the BM form,
\begin{align}
    T_{\delta\vec{G}} = \sum_{j=-1,0,1}\delta_{\delta\vec{G}+\vec{A}_j,0} \left[w_0 \rho_0 + w_1 \begin{pmatrix} 0 & \omega^j \label{FormOfT} \\  \omega^{-j} & 0 \end{pmatrix} \right], \\ \omega = e^{i \frac{2\pi}{3}}, \quad \vec{A}_0 =0, \quad \vec{A}_1 = \vec{G}_1, \quad \vec{A}_2 = \vec{G}_1 + \vec{G}_2. 
\end{align}
Note that $T_{\delta\vec{G}}^\dagger = T_{\delta\vec{G}}^\pdagger$ and $\rho_x T_{\delta\vec{G}} \rho_x = T^*_{\delta\vec{G}}$.
Furthermore, the term induced by the displacement field, $D$, is given by
\begin{equation}
    \left(h^{(D)}_{\vec{k}}\right)_{\rho,\ell,\vec{G};\rho',\ell',\vec{G}'} = -D \delta_{\rho,\rho'}\delta_{\vec{G},\vec{G}'} \begin{pmatrix} 0 & 0 & 1 \\ 0 & 0 & 0 \\ 1 & 0 & 0 \end{pmatrix}_{\ell,\ell'},
\end{equation}
which breaks the mirror symmetry, and therefore couples the different mirror-eigenvalue sectors.

The final ingredient is the proximity-induced SOC. We account for induced SOC only in the graphene layer, $l=1$, which is nearest to the WSe$_2$, i.e., the top layer in Fig.~\ref{figDW}a. Making use of the known \cite{Gmitra2015,2021arXiv210806126N} form of the proximity-induced SOC in a single graphene layer, we arrive at
\begin{subequations}\begin{align}
    H^{\text{SOC}} &= \sum_{\vec{k} \in \text{MBZ}} \sum_{\rho,\rho'=A,B}  \sum_{\eta=\pm} \sum_{s,s'=\uparrow,\downarrow}  \sum_{\vec{G}\in \text{RL}} c^\dagger_{\vec{k};\rho,1,\eta,s,\vec{G}} \left(h^{\text{SOC},l=1}_{\eta}\right)_{\rho,s;\rho',s'} c^\pdagger_{\vec{k};\rho',1,\eta,s',\vec{G}}, \\
    h^{\text{SOC},l=1}_{\eta} &=\lambda_{\text{I}} s_z \eta + \lambda_{\text{R}} \left(\eta \rho_x s_y - \rho_y s_x \right) + \lambda_{\text{KM}} \eta \rho_z s_z + m \rho_z.
\end{align}\label{SpinOrbCouplTerm}\end{subequations}
The four terms describe Ising $\lambda_{\text{I}}$, Rashba $\lambda_{\text{R}}$, and a ``Kane-Mele'' $\lambda_{\text{KM}}$ types of SOC. We also include a mass term $m$, which accounts for $C_{2z}$ breaking due to the WSe$_2$ layer. Although the inclusion of $m$ and $\lambda_{\text{KM}}$ is computationally straightforward, we set $m=\lambda_{\text{KM}}=0$ in our explicit calculations below since these two SOC terms are expected to be negligible small for a significantly misaligned (cf.~Fig.~\ref{f:fab}) WSe$_2$-graphene heterostructure \cite{2021arXiv210806126N}.

Upon transforming \equref{SpinOrbCouplTerm} to the mirror eigenbasis according to \equref{TrafoToMirrorEigenbasis}, we find
\begin{equation}
    \left(h^{(\text{SOC})}_{\vec{k},\eta}\right)_{\rho,\ell,\vec{G};\rho',\ell',\vec{G}'} = \frac{\delta_{\vec{G},\vec{G}'}}{2} \begin{pmatrix} (h^{\text{SOC},l=1}_{\eta})_{\rho,\rho'} & 0 & (h^{\text{SOC},l=1}_{\eta})_{\rho,\rho'} \\ 0 & 0 & 0 \\ (h^{\text{SOC},l=1}_{\eta})_{\rho,\rho'} & 0 & (h^{\text{SOC},l=1}_{\eta})_{\rho,\rho'} \end{pmatrix}. \label{SOCTransformed}
\end{equation}
The presence of WSe$_2$ on a single side breaks the mirror reflection symmetry explicitly, and therefore induces couplings between the mirror-even and mirror-odd sectors [off-diagonal terms in \equref{SOCTransformed}]. While we take these off-diagonal terms into account in our band structure calculations, we expect that their impact is subleading for generic momenta $\vec{k}$ where the (approximately) mirror-even and mirror-odd bands for $D=0$ (at small $D$) are energetically well-separated. In this limit, the main impact of the SOC coupling onto the mirror-even, ``twisted-bilayer-graphene-like'' \cite{Christos2021tTLG}, bands is the same as that of \WSe\ on twisted-bilayer graphene (modulo rescaling of parameters); the same holds for the mirror-even, ``graphene-like'', sector.

This completes our definition of all terms in the continuum-model Hamiltonian $H_0^{\text{Full}}$ in \equref{FullContinuumModel}.

\vspace{1em}

\textbf{Parameters and twist-angle dependence.} 
The noninteracting model \eqref{FullContinuumModel} predicts a magic angle near $\theta\sim1.55^{\circ}$, for the standard set of parameters \cite{Bistritzer2011}: $v_F/a=2.7\sqrt{3}/2 \times 10^3$ meV, $w_0=w_1=110$ meV, and zero SOC. 
The resulting bandwidth for a twist angle of $\theta\sim1.37^{\circ}$ is much larger than what we observe experimentally for this angle [see Fig.~\ref{fig1}d in the main text]. We have also checked that adding SOC does not change this conclusion.

To account for this difference within the noninteracting model, we allow for variations in the sublattice diagonal and off-diagonal interlayer hopping strengths, $w_0$ and $w_1$ respectively. This is motivated by the observation, in the absence of WSe$_2$ layer, that $w_0<w_1$ takes into account corrugation effects \cite{NguyenKoshino2017,PhysRevResearch.1.013001,2021arXiv211111060L}.  
With this in mind, Fig.~\ref{f:theoryplots}a shows the twist-angle dependence of the moir\'e bandwidth for two cases: (i) $w_0=0.875 w_1, \ w_1=w=110$ meV, and (ii) $w_0=w_1=0.86 w, \ w=110$ meV. A qualitative difference is seen; in case (i) there is a {\it magic dip} at $\theta \approx 1.37^{\circ}$, while for (ii) there is a direct shift of the {\it magic angle} down to $\theta\approx1.37^{\circ}$ (as expected since scaling the inter-layer coupling down is equivalent to decreasing the magic angle \cite{Bistritzer2011}). Further accounting for SOC on top of this broadens the bandwidth, and makes both cases consistent with the experimentally observed bandwidth of $\sim10\,\textrm{meV}$, see Fig.~\ref{fig1}d of the main text; in that plot, we used $\lambda_{\text{I}}=\lambda_{\text{R}}=10\,$meV where $\lambda_{\text{I}}=\lambda_{\text{R}}$ is motivated by the fact that these two SOC terms have been estimated \cite{2021arXiv210806126N} to be roughly the same for a relative of twist angle of $\approx 20^{\circ}$ between graphene and \WSe\, see Fig.~\ref{f:fab}, and the absolute strength is chosen to approximately reproduce Fig.~\ref{fig1}c of the main text. All results presented below and in the main text will take the inter-layer hopping parameters from case (i) and $\lambda_{\text{I}}=\lambda_{\text{R}}=10\,$meV.

\vspace{1em}

\textbf{Density of states.}
Computing the density of states of the noninteracting Hamiltonian, taking $w_0=0.875 w_1, \ w_1=w=110$ meV and $\lambda_{\text{I}}=\lambda_{\text{R}}=10$meV, Figure \ref{f:theoryplots}b and c demonstrate two keys features that are seen experimentally [namely Fig.~\ref{figDW}a and b of the main text]: (i) the van Hove singularity (vHs) becomes more pronounced at finite $D$, although this enhancement is more subtle than seen experimentally, and (ii) that there is an asymmetry of $D>0$ and $D<0$ at fixed $\nu$, but approximate invariance under $(D,\nu) \rightarrow (-D,-\nu)$.

\begin{figure}[h]
{\includegraphics[width=0.85\textwidth,clip]{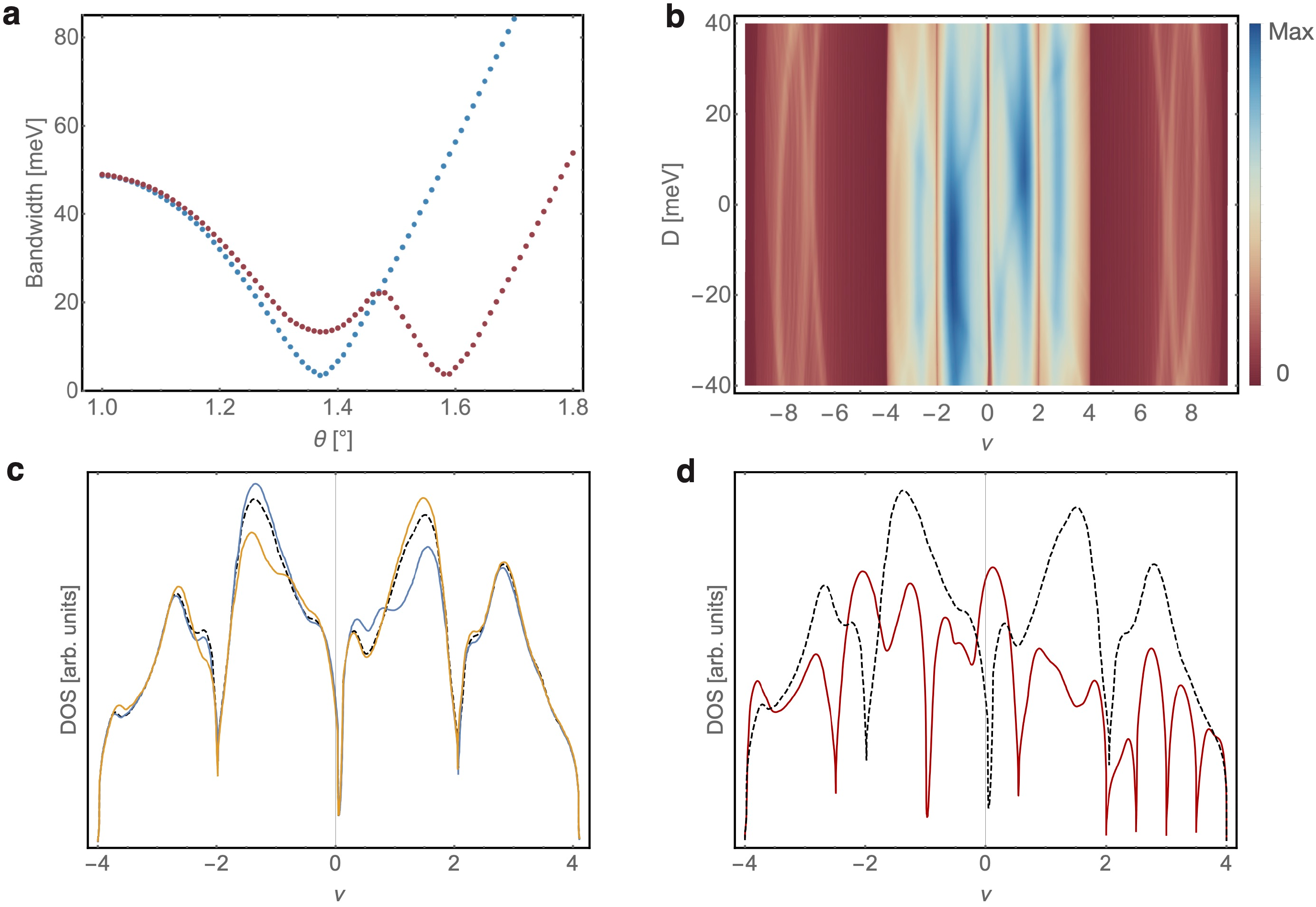}}
\caption{ \label{figmodel} (a) Bandwidth vs.~twist angle, where blue and red points correspond to $w_0=w_1=0.86w$, $w=110$ meV and $w_0=0.875w, w_1=w$, $w=110$ meV, respectively, and SOC is set to zero. Note the additional dip at $\theta\approx 1.37^{\circ}$ for the red points. In (b), (c) and (d) we take $w_0=0.875w, w_1=w$, $w=110$ meV, and $\lambda_{\text{I}}=\lambda_{\text{R}}=10$ meV. (b) Density of states as a function of $\nu$ and $D$. (c) Black dashed, solid blue and solid orange correspond to $D=\{0, 10, -10\}$ meV, respectively.  An approximate symmetry $(D,\nu)\to(-D,-\nu)$ is seen. Moreover, the van Hove peak becomes slightly more prominent for $D=\pm 10$ meV relative to $D=0$. (d) Red curve includes a DW order parameter $\Phi_0=4$meV, according to \equref{HDW}, while the black dashed line is the reference density of states at $\Phi_0=0$. }
\label{f:theoryplots}
\end{figure}

\subsection{SI 3: Density-wave order parameters}

\noindent \textbf{Possible CDW order parameters.} Let us begin our theoretical discussion of possible DW phases with conventional charge density wave (CDW) order. By CDW we mean phases where moir\'e translational symmetry is broken by a spatial modulation of the electronic density while all flavor symmetries (in our case valley, but without SOC also spin) are preserved. 
Restricting the analysis to order parameters with momentum transfer given by any of the three M points of the MBZ, $\vec{M}_j$, $j=1,2,3$, the presence of CDW order can be described by a coupling of the form
\begin{equation}
    H_\phi = \sum_{\vec{R}\in\text{ML}} \sum_{\eta,s,\alpha} \sum_{j=1}^3 \widetilde{\phi}_j \, e^{i \vec{M}_j\vec{R}} c^\dagger_{\vec{R},\alpha,\eta,s} c^\pdagger_{\vec{R},\alpha,\eta,s} + \text{H.c.} = \sum_{\vec{R}\in\text{ML}} \sum_{\eta,s,\alpha} \sum_{j=1}^3 \phi_j \cos(\vec{M}_j\vec{R})\, c^\dagger_{\vec{R},\alpha,\eta,s} c^\pdagger_{\vec{R},\alpha,\eta,s}. \label{NormalCDWAnsatz}
\end{equation}
Here $c^\dagger_{\vec{R},\alpha,\eta,s}$ creates an electron with spin $s$, in valley $\eta$, and in Wannier state $\alpha$ in unit cell $\vec{R}$ associated with a set of low-energy bands of interest. In the second equality, we used that $e^{i \vec{M}_j\vec{R}} = e^{-i \vec{M}_j\vec{R}}$ allowing us to introduce the real-valued, three-component DW order parameter $\phi_j = \tilde{\phi}_j + \tilde{\phi}_j^* \in \mathbb{R}$.

Upon noting that the action of $C_{3z}$ is $\phi_j \rightarrow \phi_{(j+1 \text{mod}3)}$ and that of an elementary translation by a primitive vector $\vec{a}_{j'}$ is given by $\phi_{j'} \rightarrow \phi_{j'}$, $\phi_{j\neq j'} \rightarrow -\phi_{j}$, the free-energy up to quartic order in $\phi_j$ can only have the form
\begin{equation}
    \mathcal{F} \sim a(T)\, \sum_j \phi_j^2 + b \, \phi_1\phi_2\phi_3 + c_1\, \Bigr(\sum_j \phi_j^2\Bigr)^2 + c_2\, \left[ (\phi_1\phi_2)^2 + (\phi_2\phi_3)^2 + (\phi_3\phi_1)^2 \right], \label{FreeEnergyExpansion}
\end{equation}
where $a$, $b$, $c_1$, and $c_2$ are real-valued, phenomenological parameters.
To discuss the resulting phases, let us first consider $b\rightarrow 0$. It is easy to see that there are only two possible minima: if $c_2>0$, we find
\begin{equation}
    \phi_1 = \Phi_0, \quad \phi_2=\phi_3=0,
\end{equation}
where $\Phi_0>0$ without loss of generality, and symmetry-related configurations. This is the 2-unit-cell state in the upper panel of Fig.~\ref{figDW}d. If $c_2<0$, we instead find
\begin{equation}
    \phi_1 = \phi_2=\phi_3= -\text{sign}(b)\,\Phi_0, \qquad \Phi_0 > 0 , \label{FourUnitCellState}
\end{equation}
as well as its symmetry-related states; this corresponds to the 4-unit-cell state shown in the lower panel, left part, of Fig.~\ref{figDW}d. While finite values of $b$ in \equref{FreeEnergyExpansion} do not lead to new phases as $\phi_1\phi_2\phi_3$ is extremized if $|\phi_1| = |\phi_2| = |\phi_3|$, $b\neq 0$ has crucial consequences for the nature of the thermal phase transition: as is generically expected to be the case, let us assume that the temperature dependence of $b$, $c_1$, and $c_2$ can be neglected near the critical temperature $T_c$, where $a(T)$ changes sign. We then see that, irrespective of the sign of $c_2$, the system will always first enter the 4-uni-cell state in \equref{FourUnitCellState} right below $T_c$. If $c_2<0$, the system will stay in this phase at lower temperature [at least, as long as the expansion (\ref{FreeEnergyExpansion}) is valid]. However, in the case $c_2>0$, there will be a first order transition into the 2-unit-cell state at some temperature smaller than $T_c$. While these two temperatures can in principle be very close and, hence, might be hard to resolve, we do not see any sign of first-order transitions in the DW states, which provides further evidence in favor of the 4-uni-cell state in the lower left panel of Fig.~\ref{figDW}d.

\vspace{1em}

\noindent \textbf{Band reconstruction.} 
We next consider the influence of the $4$-unit-cell CDW order, associated with the minimum in \equref{FourUnitCellState}, on the band structure. 
The corresponding impact on the moir\'e flat bands (i.e. the four bands, per valley $\eta$, in the vicinity of charge neutrality) is captured by the Hamiltonian,
\begin{align}
\label{HDW}
H^{\text{DW}}&=\sum_{\bm k}\sum_{\eta=\pm}\sum_{n=1}^4 \left( d^\dag_{\eta,\bm k,n}d^\pdagger_{\eta,\bm k,n}\varepsilon_{\eta,\bm k,n} + \Phi_0  \sum_j d^\dag_{\eta,\bm k,n}d^\pdagger_{\eta,\bm k\pm\bm M_j,n} \right),
\end{align} 
where $d^\dag_{\eta,\bm k,n}$ are creation operators in the band basis of the noninteracting Hamiltonian \eqref{FullContinuumModel}, with eigenenergies $\varepsilon_{\eta,\bm k,n} $. Here $n$ is the band index, and momentum $\bm k$ is restricted to the reduced Brillouin zone associated with the broken translational symmetry of the CDW order.  

Figure \ref{f:theoryplots}d plots the density of states found from \eqref{HDW}, with $\Phi_0=4$ meV, which provides a simple demonstration that the presence of the DW acts to split the moir\'e bands, generating many additional van Hove singularities.  This provides a natural explanation for the additional features seen in the Hall number at low temperature, see Fig.~\ref{figSC}e, Fig.~\ref{figDW}d-f and Fig.~\ref{HallT}c.

\vspace{1em}

\noindent \textbf{Intervalley coherent DWs.} As alluded to in the main text, the measurements are not only consistent with simple CDW phases as defined above, but also with more exotic DW phases, as we will illustrate next. As spin-rotation invariance is already broken by the induced SOC, let us focus on states where real-space translations are intertwined with U(1)$_v$, the group of independent U(1) phase transformations in the two valleys. In that case, \equref{NormalCDWAnsatz} is replaced by
\begin{equation}
    H_\Phi = \sum_{\vec{R}\in\text{ML}} \sum_{\eta,s,\alpha} \sum_{j=1}^3\sum_{j'=1,2} \Phi_{j,j'} \cos(\vec{M}_j\vec{R})\, c^\dagger_{\vec{R},\alpha,\eta,s} (\eta_{j'})_{\eta,\eta'} c^\pdagger_{\vec{R},\alpha,\eta',s}.
\end{equation}
Here, the underlying order parameter is a real-valued $3\times 2$ matrix $\Phi$ or, equivalently, three $2$-component real vectors $\vec{\varphi}_j$ with $(\vec{\varphi}_j)_{j'} = \Phi_{j,j'}$. Under U(1)$_v$, these vectors transform as $\vec{\varphi}_j \rightarrow e^{i\eta_y \varphi} \vec{\varphi}_j$, while $\vec{\varphi}_j$ transform the same way as $\phi_j$ in \equref{NormalCDWAnsatz} under $C_{3z}$ and translations. One difference is that they transform non-trivially under $C_{2z}$, $\vec{\varphi}_j \rightarrow -\eta_z \vec{\varphi}_j$.

With these constraints, it is easy to show that the most general free-energy expansion reads as
\begin{equation}
    \mathcal{F} \sim a(T) \sum_j \vec{\varphi}_j^2 + c_1\, \Bigr(\sum_j \vec{\varphi}_j^2\Bigr)^2 + c_2\, \sum_{j_1\neq j_2} (\vec{\varphi}_{j_1} \cdot \vec{\varphi}_{j_2})^2 + c_3\, \sum_{j_1\neq j_2} \vec{\varphi}_{j_1}^2 \vec{\varphi}_{j_2}^2. \label{FreeEnergyIVC}
\end{equation}
Note that, as opposed to \equref{FreeEnergyExpansion}, no third-order term is possible, which is a consequence of U(1)$_v$. The free energy in \equref{FreeEnergyIVC} allows for the following four phases: two of them,
\begin{equation}
    \vec{\varphi}_{1} = \vec{\varphi}, \quad \vec{\varphi}_{2,3} = 0
\end{equation}
and 
\begin{equation}
    \vec{\varphi}_{1} = \vec{\varphi}, \quad \vec{\varphi}_{2} = \hat{\vec{z}} \times  \vec{\varphi}, \quad \vec{\varphi}_{3} = 0.
\end{equation}
are nematic, i.e., break $C_{3z}$ symmetry and are, hence, less natural candidates than the $C_{3z}$ preserving phases with
\begin{equation}
    \vec{\varphi}_{j} = \vec{\varphi} \label{SimplestIVCDWOrder}
\end{equation}
and
\begin{equation}
    \vec{\varphi}_{j} = (C_{3z})^{j-1}\vec{\varphi}.
\end{equation}
respectively.

For instance, the state defined by \equref{SimplestIVCDWOrder} corresponds to a coupling (choosing $\vec{\varphi}=\vec{e}_x$ for concreteness)
\begin{equation}
    H_\Phi = \sum_{\vec{R}\in\text{ML}} \sum_{\eta,s,\alpha}  \Phi(\vec{R})\, c^\dagger_{\vec{R},\alpha,\eta,s} (\eta_{1})_{\eta,\eta'} c^\pdagger_{\vec{R},\alpha,\eta',s}, \qquad \Phi(\vec{R}) = \sum_{j=1}^3 \cos(\vec{M}_j\vec{R}).
\end{equation}
The impact of DW order of this form on the band energies will be similar to that of a simple CDW as in \equref{HDW}. However, since the resulting mini bands will be time-reversal symmetric superpositions of two valleys, sequentially filling these mini bands will always lead to DW states with vanishing Chern number, irrespective of the Chern numbers of the SOC coupled bands (cf.~Fig.~\ref{figCI}b of the main text, where we show one possible distribution of these Chern numbers). Further measurements are required, though, to be able to determine which of the different DW states is realized in the system.

\subsection{SI 4: Transport behavior in the small twist angle regimes}

\newpage

\begin{figure*}[h]
{\includegraphics[width=0.57\textwidth,clip]{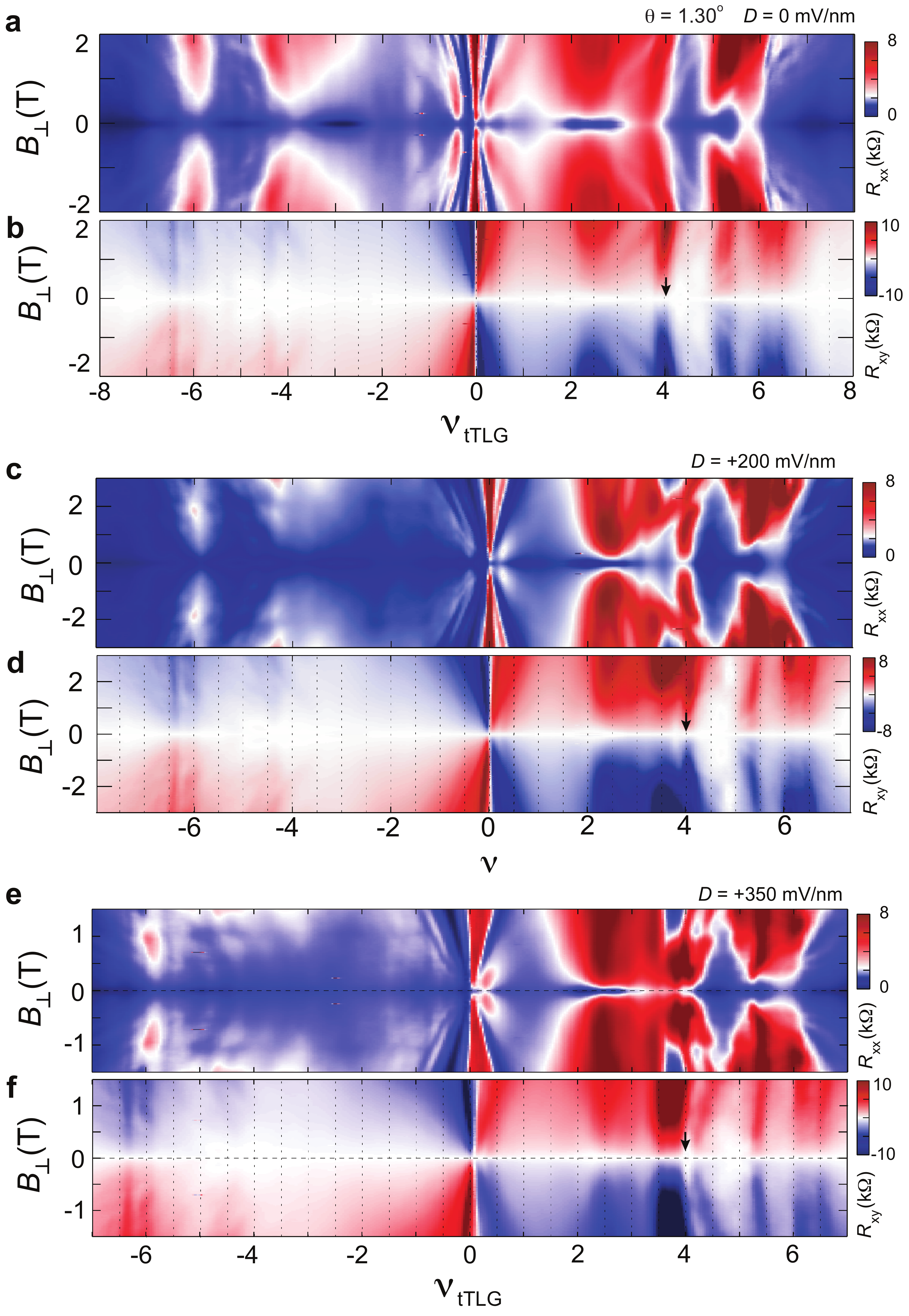}
\caption{\label{figB}\textbf{Magnetotransport data measured from Sample B} Sample B consists of a tTLG with twist angle of $\theta = 1.30^{\circ}$. There is no atomic interface between tTLG and \WSe. (a), (c) and (e) plot longitudinal resistance as a function of filling fraction $\nu_{tTLG}$ and magnetic field $B$. (b), (d) and (f) show Hall resistance as a function of filling fraction $\nu_{tTLG}$ and magnetic field $B$ measured. (a-b), (c-d) and (e-f) are measured with $D=0$, $D = 200$ mV/nm and $D = 350$ mV/nm, respectively. All measurements are performed at $T = 20$ mK. The vertical arrow in (b) and (d) point towards robust incompressible state emerging from $\nu_{tTLG}=+4$, which is used to determine the transition from the primary to remote moir\'e band in the electron-doped band. On the hole-doping side, such a transition is much weaker. At $D = 0$, $\nu_{tTLG}=-4$ can be associated with features in the longitudinal channel. However, no clear transport response mark this transition at $D=+200$ and $+300$ mV/nm. This is consistent with the band structure calculated using the continuum model, where the primary and remote moir\'e bands are connected at the $\Gamma$ point. Despite the absence of a tTLG/\WSe\ interface, characteristic phenomenology of the small twist angle regime is observed: incompressible states exhibit density modulation with $1/4$ and $1/2$ moir\'e fillings. Dotted vertical lines in (b), (d) and (f) mark moir\'e fillings of  $N/2$, where $N$ takes integer values. Superconductivity is observed in the density range of $2 < |\nu_{tTLG}| < 4$ near $B = 0$.  Landau levels associated with the Dirac band is suppressed.  }}
\end{figure*}

\begin{figure*}[h]
{\includegraphics[width=0.7\textwidth,clip]{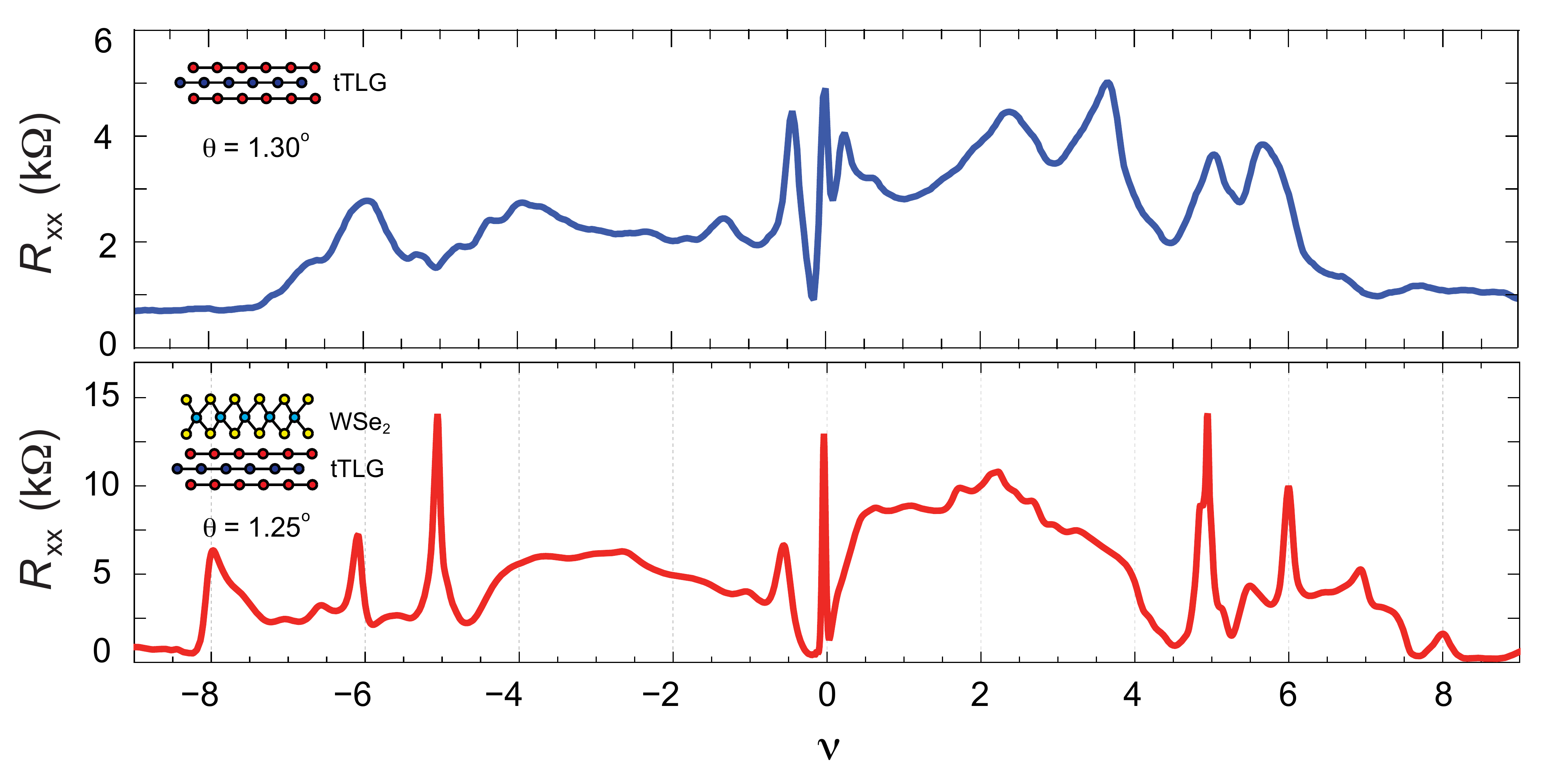}
\caption{\label{f:tTLG}\textbf{tTLG in the small twist angle regime} Longitudinal resistance \Rxx\ as a function of moir\'e filling $\nu_{tTLG}$ measured from sample A (bottom panel) and B (top panel). Measurements are performed in the presence of a small out-of-plane magnetic field to suppress superconductivity. Sample A consists of a \WSe\/tTLG heterostructure with a twist angle of $\theta = 1.25^{\circ}$. Sample B consists of just tTLG \textbf{without} \WSe\ at a twist angle of $\theta = 1.30^{\circ}$. Resistive features are sharper in sample A, likely indicative of a more uniform sample. Both samples exhibit resistance peaks in the remote bands $|\nu_{tTLG}|>4$.  }}
\end{figure*}

\newpage
\begin{figure*}[h]
{\includegraphics[width=0.9\textwidth,clip]{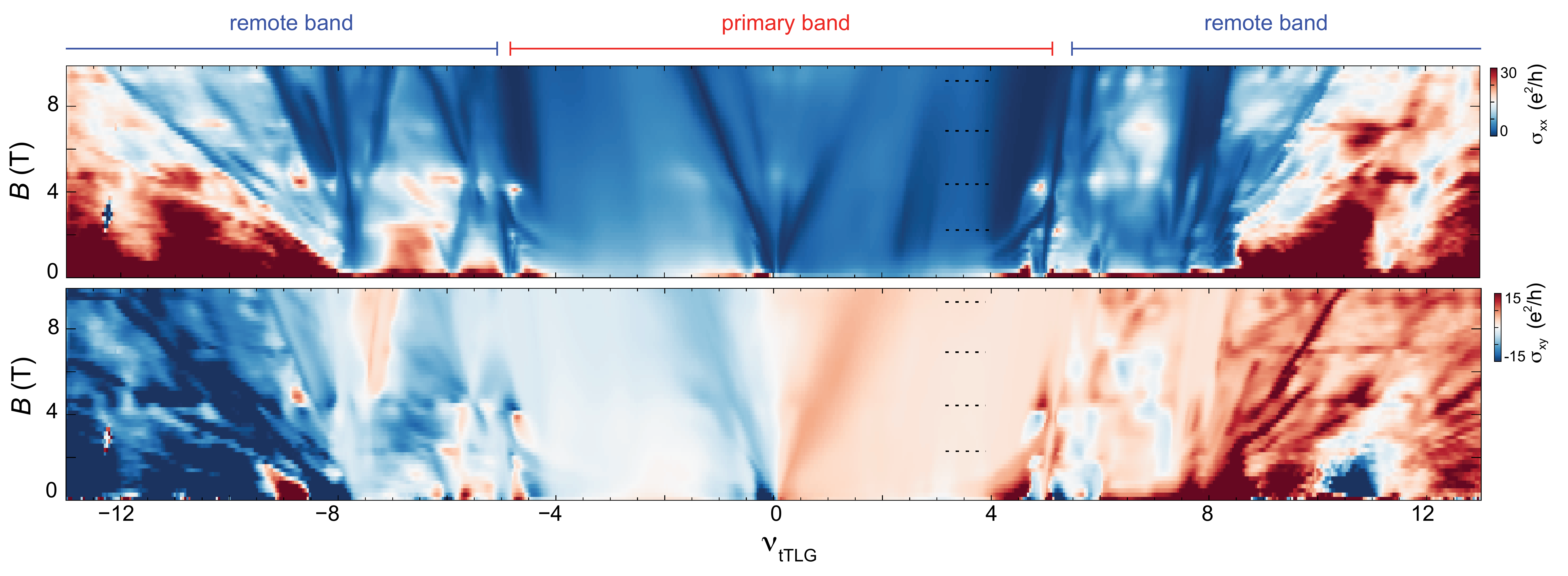}
\caption{\label{f:spectrum}\textbf{tTLG in the small twist angle regime}  Longitudinal conductance $\sigma_{xx}$ (top panel) and Hall conductance $\sigma_{xy}$ (bottom panel) as a function of $\nu_{tTLG}$ and $B$, measured at $D =-300$ mV/nm and $T=1$ K. The horizontal dashed lines correspond to high-symmetry points in the Hofstadter spectrum.  }}
\end{figure*}

\begin{figure*}[h]
{\includegraphics[width=0.8\textwidth,clip]{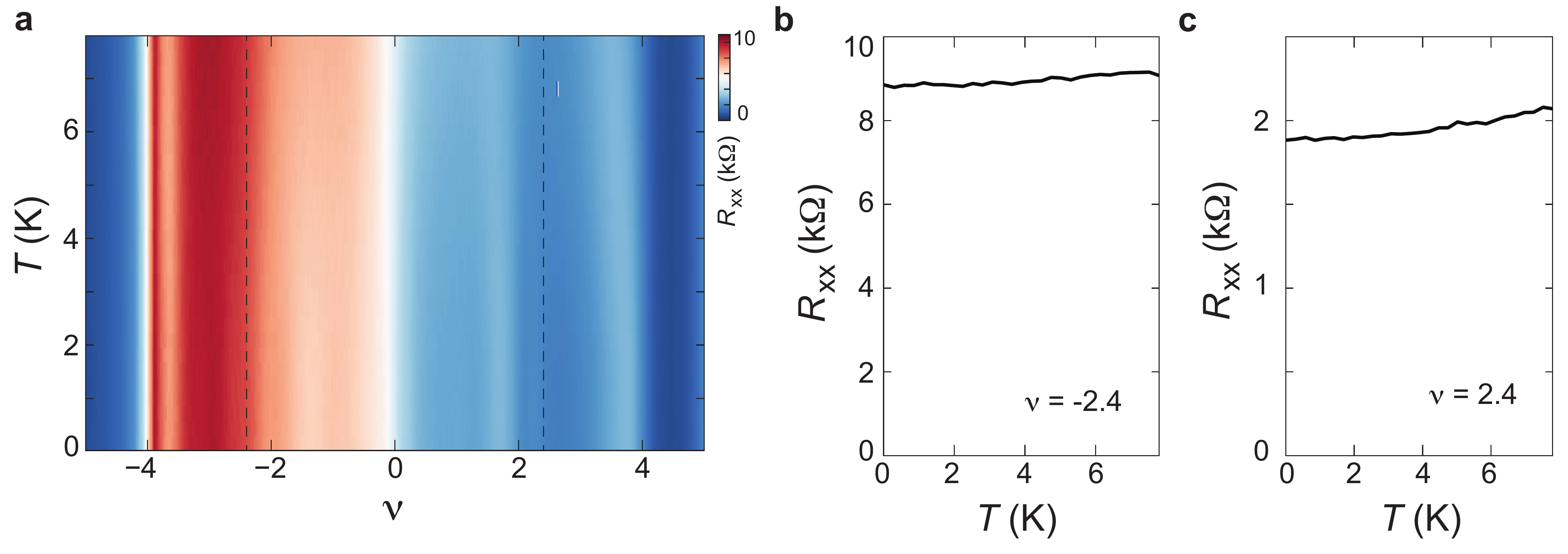}
\caption{\label{f:E}\textbf{The absence of superconductivity in  sample E} (a) Longitudinal Resistance $R_{xx}$ as a function of filling fraction $\nu_{tTLG}$ and temperature $T$ measured from sample E at $\theta=1.38^{\circ}$. (b-c) \Rxx\ as a function of $T$ measured at constant carrier density of (b) $\nu=-2.4$ and (c) $+2.4$. No superconducting behavior is observed in sample E.  }}
\end{figure*}

\begin{figure*}[h]
{\includegraphics[width=0.8\textwidth,clip]{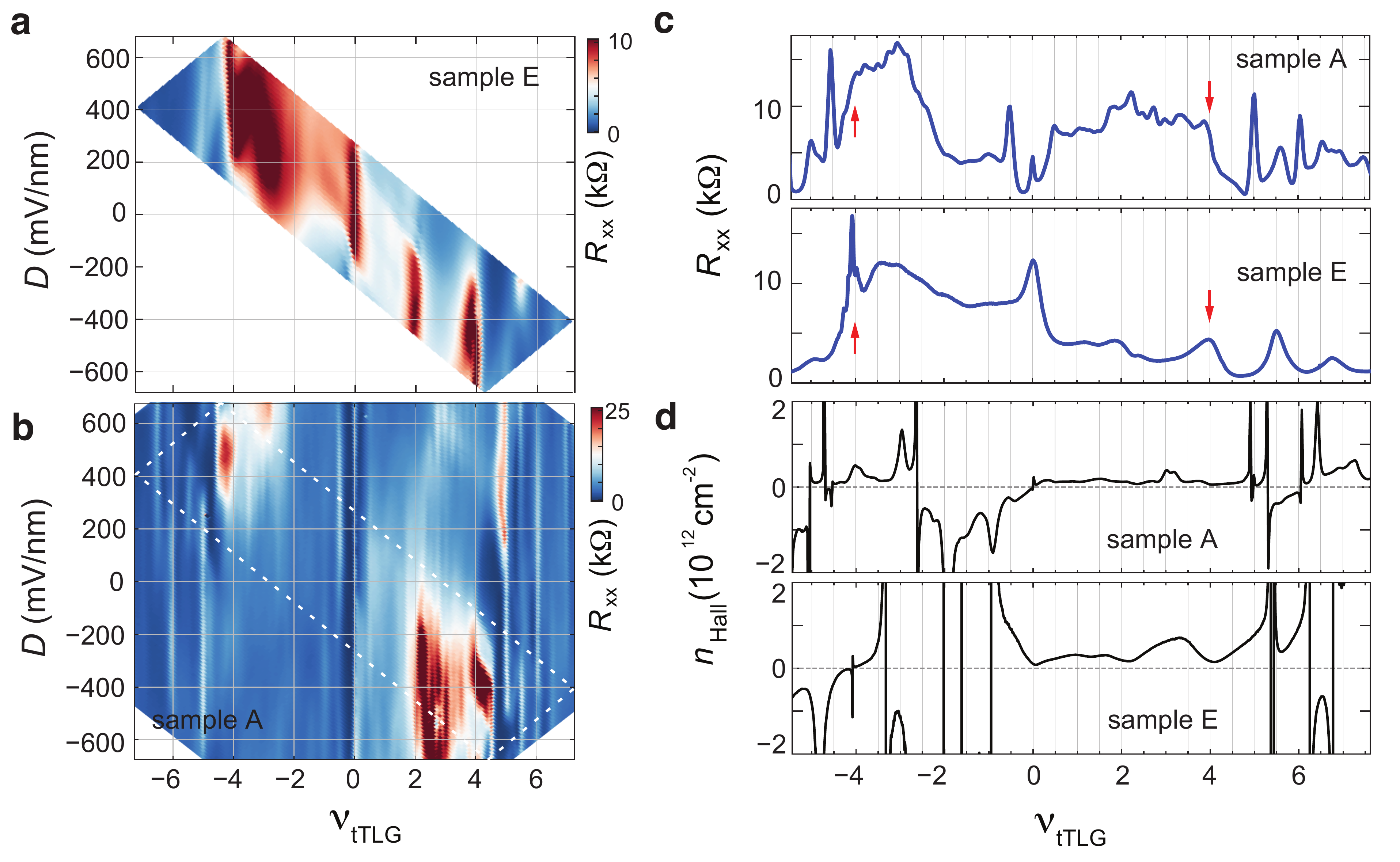}
\caption{\label{f:AB}\textbf{Comparison between sample E and sample A} Longitudinal Resistance $R_{xx}$ as a function of filling fraction $\nu_{tTLG}$ and displacement field $D$ for (a) sample E (b) sample A. The white dashed line in (b) marks the boundary of the area mapped in (a). (a) is measured at $B = 2$ T, where the resistance peak at $\nu_{tTLG} = +2$ is enhanced by the application of the magnetic field.  (c) $R_{xx}$ and (d) $n_H$ as a function of $\nu_{tTLG}$ measured from (top panel) sample A and (bottom panel) sample B. Measurements are performed at $T=20$ mK and $B=0.4$ T. The red arrows in (c) mark the position of $\nu = \pm4$, which is defined by the most robust incompressible state in the quantum Hall effect regime, as shown in Fig.~\ref{figfanA}. Both sample A and sample E feature resistance peaks in the remote band. Hall density also exhibits resets and vHSs in the remote band. }}
\end{figure*}



\subsection{SI 5: determining twist angle and moir\'e filling in sample A}

\begin{figure*}[h]
{\includegraphics[width=0.45\textwidth,clip]{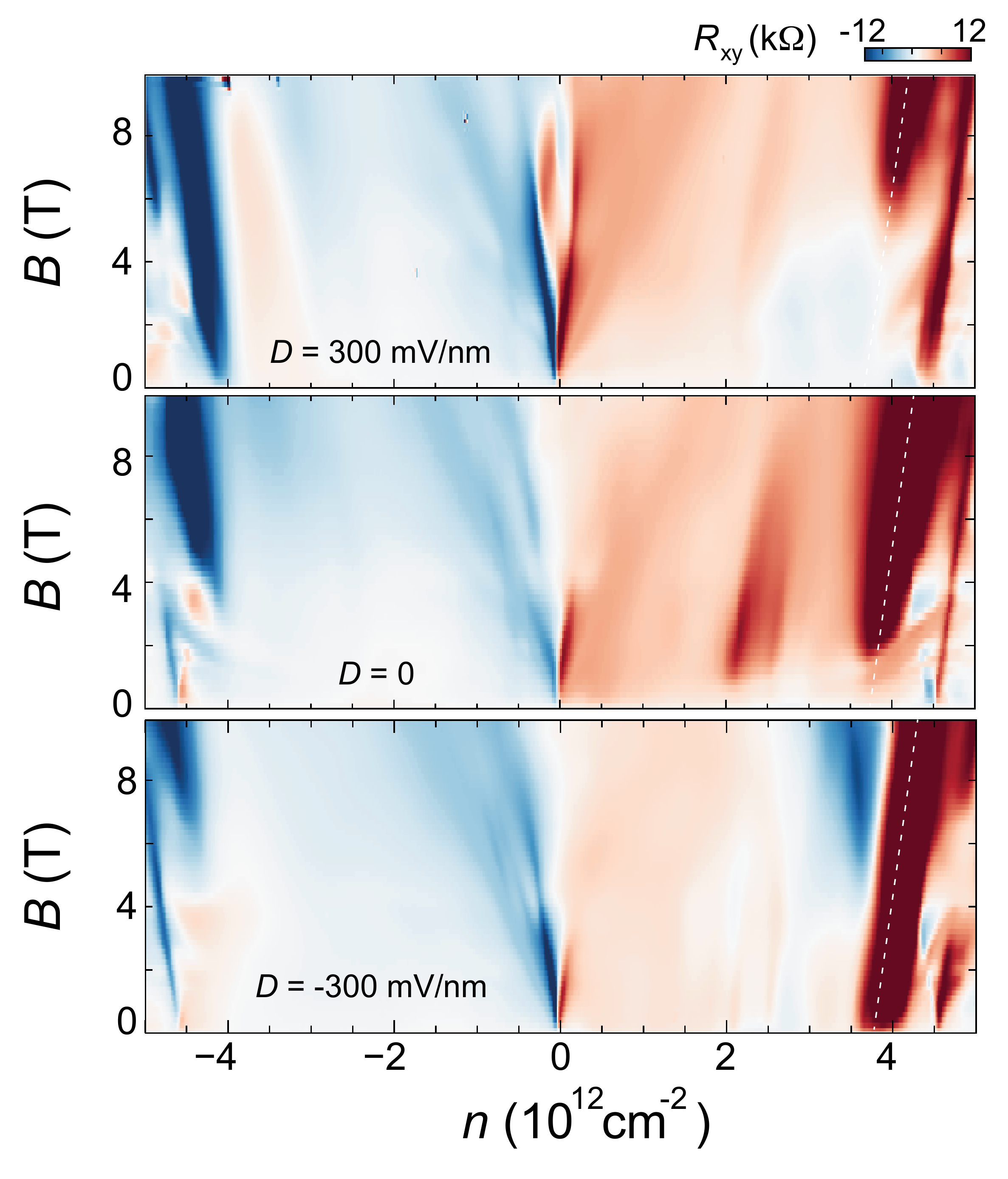}
\caption{\label{figfanA}\textbf{Incompressible states in the quantum Hall effect regime} Hall Resistance $R_{xy}$ as a function of magnetic field $B$ and carrier density $n$ at $T=20$ mK. The measurements in the top, middle and bottom panels are performed at $D=+300$, $0$ and $-300$ mV/nm, respectively.}}
\end{figure*}

In sample A,  the most robust incompressible state in the quantum Hall effect regime  are marked by the white dashed lines in Fig.~\ref{figfanA}. We identify the zero field position of the white dashed lines as $\nu_{tTLG}=+4$, which allows us to determine the twist angle to be $\theta = 1.25^{\circ}$.. 
At $D= 300$ mV/nm, the incompressible state from $\nu=+4$ disappears at $B < 5$ T, which is indicative of complex band crossings in the energy band structure. This is consistent with the calculated band structure in Fig.~\ref{fig1}k. 

\begin{figure*}[h]
{\includegraphics[width=0.8\textwidth,clip]{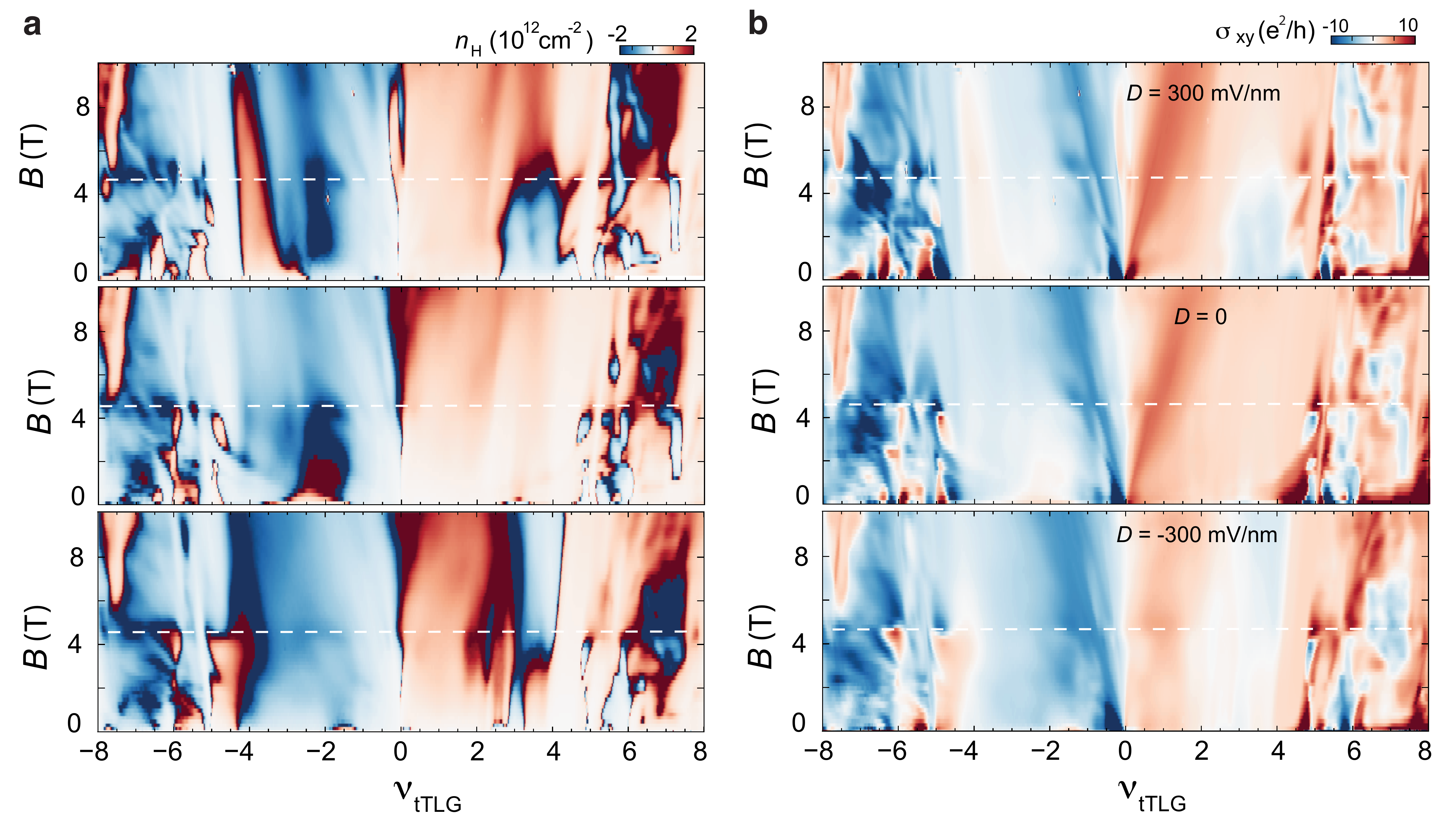}
\caption{\label{f:sampleA2}\textbf{High symmetry point in the Hofstadter spectrum} (a) Hall density $n_H$ and (b) Hall conductance $\sigma_{xy}$ as a function of $B$ and filling fraction $\nu_{TLG}$ at $T=20$ mK. The measurements in the top, middle and bottom panels are performed at $D=+300$, $0$ and $-300$ mV/nm, respectively. The white dashed lines indicate the magnetic field in which there is one magnetic flux per 8 moir\`e unit cells}}
\end{figure*}

Notably, the twist angle can be independently determined based on the magnetic field value of the  high symmetry point in the Hofstadter spectrum, since the horizontal dashed lines in Fig.~\ref{figCI} correspond to integer values of $\phi_0/\phi$. Assigning the high symmetry point around $B= 4.85$ T as $\phi_0/\phi = 4$ gives rise to a twist angle of $\theta = 1.22^{\circ}$, which is consistent with the twist angle determined from the Landau fan.
Fig.~\ref{f:sampleA2}b plots Hall conductance $\sigma_{xy}$ as a function of $B$ and $\nu_{tTLG}$ at different $D$. The high symmetry point of the Hofstadter spectrum, which is marked by the horizontal white dashed lines, is independent of $D$, suggesting that the moir\'e wavelength is the same between top/middle and middle/bottom graphene layers. In another word, the tTLG sample has a A-tw-A stacking order ~\cite{Park2021flavour}.

\begin{figure*}[h]
{\includegraphics[width=0.85\textwidth,clip]{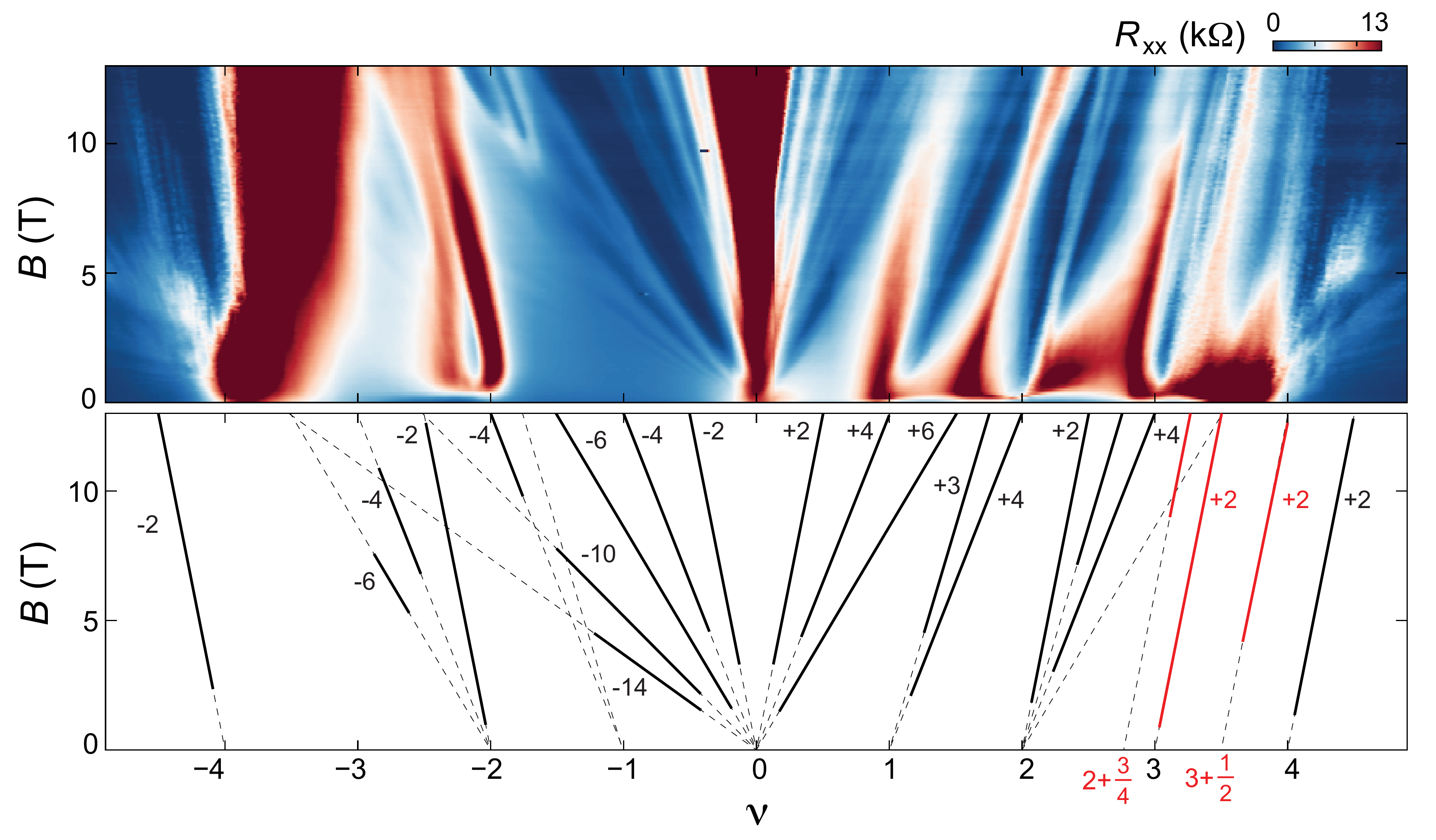}
\caption{\label{f:sampleC}\textbf{Magnetotransport data measured from sample C} (top panel) Longitudinal resistance $R_{xx}$ as a function of magnetic field $B$ and moir\'e filling $\nu$ measured at $T=20$ mK and $D =0$ in sample C, which has a twist angle of $\theta = 1.5^{\circ}$. (bottom panel) Schematic Landau fan capturing the most prominent incompressible states in the Landau fan. Around the charge neutrality point, the Landau fan exhibits similar behavior compared to magic-angle tTLG without proximity ~\cite{Liu2021DtTLG}. The Landau fan emanating from the CNP has a main sequence of $\pm2$, $\pm6$, $\pm10$ ... Extra Landau fans are observed emerging from every integer filling at $\nu=\pm1$, $\pm2$ ... This demonstrates that Fermi surface reconstruction at integer filling plays a dominating role. In the filling fraction range $2 < \nu <4$, we observe a series of incompressible states with fractional intercept of $s=2+3/4$ and $3+1/2$. The fractional value in $s$ is similar to the behavior in sample A and B, suggesting that DW instability with $1/2$ and $1/4$ density modulation is universal for tTLG samples. Near the magic angle, flavor polarization at integer fillings play a dominating role and DW instability has a weaker energy scale. In the small twist angle regime, as short range Coulomb and flavor polarization are suppressed, the influence of long range Coulomb and DW instability become more prominent. }}
\end{figure*}

\begin{figure*}[h]
{\includegraphics[width=0.4\textwidth,clip]{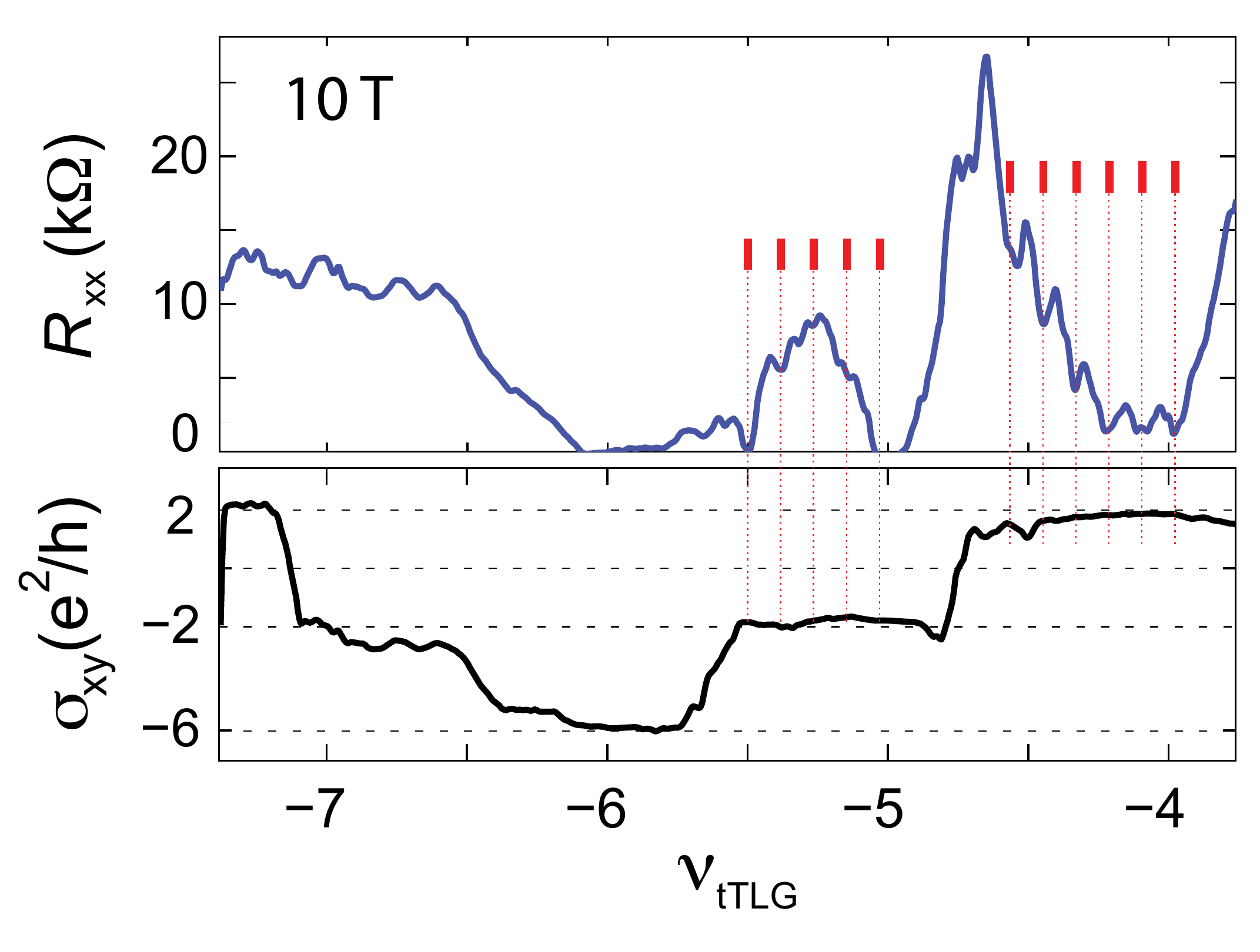}
\caption{\label{f:linecut}\textbf{Density modulation of 1/8 moir\'e filling.} \Rxx\ (top panel) and $\sigma_{xy}$ (bottom panel) as a function of $\nu_{tTLG}$ measured at $D=500$ mV/nm, $B=10$ T and $T = 20$ mK. Over most of the density range in the remote band, incompressible states have Hall resistance plateau with quantization value of $\pm2$, $\pm6$ ... The sequence of incompressible states reveals that fermi surface resulting from DW instability resembles that of Dirac cones. At the same time, they inherit the 4-fold degeneracy. One natural interpretation is that the interaction-reduced band reconstruction leads to revival of the Dirac cones in the remote bands at those fillings, possibly similar to that in twisted bilayer graphene~\cite{Zondiner2020cascade}, while retaining the valley and spin flavor symmetry.  }}
\end{figure*}

\begin{figure*}[h]
\includegraphics[width=0.9\linewidth]{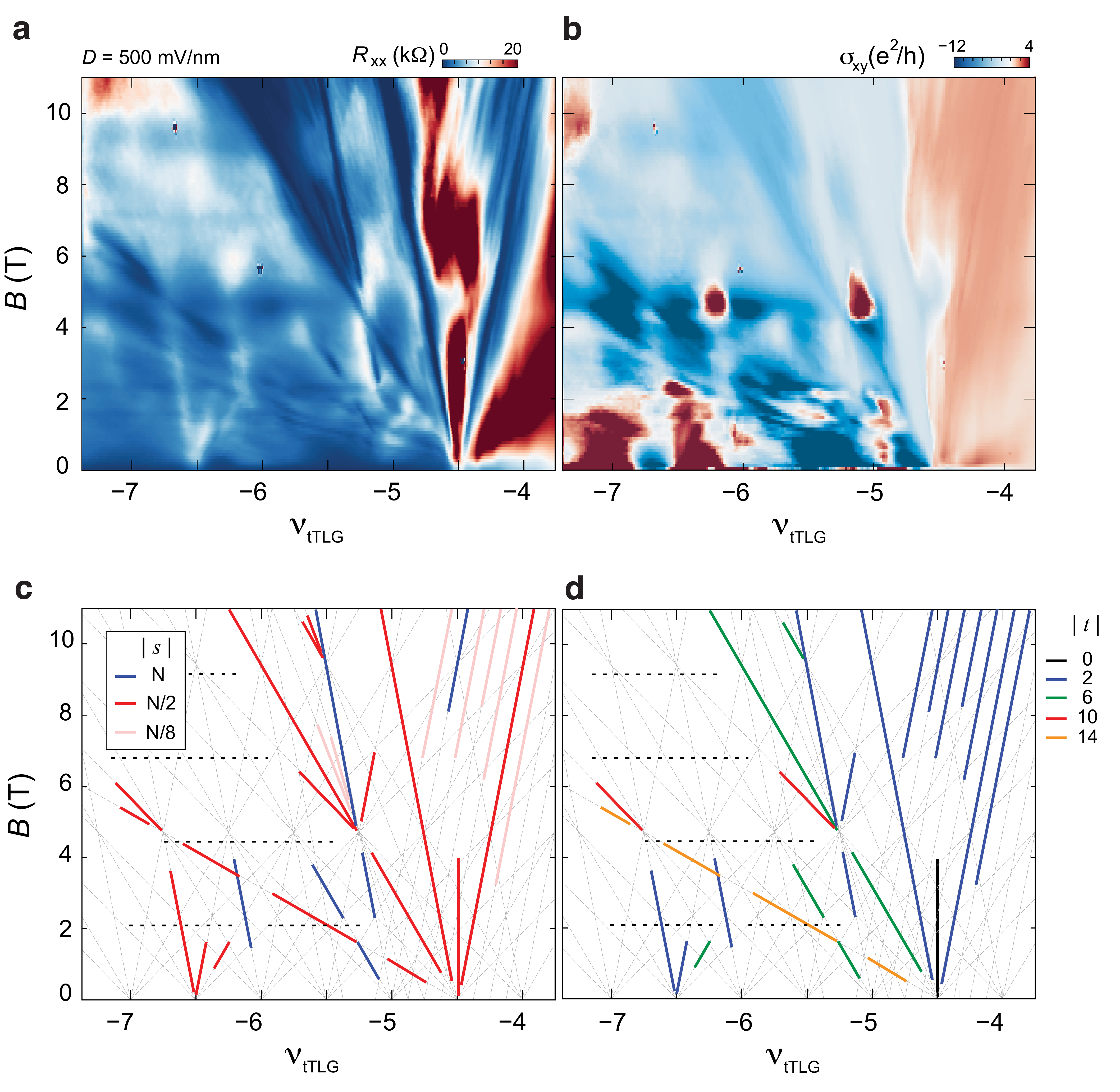}
\caption{\label{figSBCI} {\bf{SBCI in the remote band.}} $\nu_{tTLG}-B$ map of (a) \Rxx\ and (b) $\sigma_{xy}$ in the remote band measured at  $D = 500$ mV/nm and $T = 20$ mK. (c-d) Schematics of main features in (a) and (b). (c) Incompressible states are color coded according to their intercept $|s|$. Blue, red and pink solid lines denote incompressible states with intercept $s$ with integer, half integer and $N/8$ values.  (d) Incompressible states are color coded according to their slope $t$. Black, blue, green, red and orange solid lines denote incompressible states with slope $t=0$, $\pm2$, $\pm6$, $\pm10$ and $\pm14$, respectively. }
\end{figure*}

\begin{figure*}
{\includegraphics[width=0.9\linewidth]{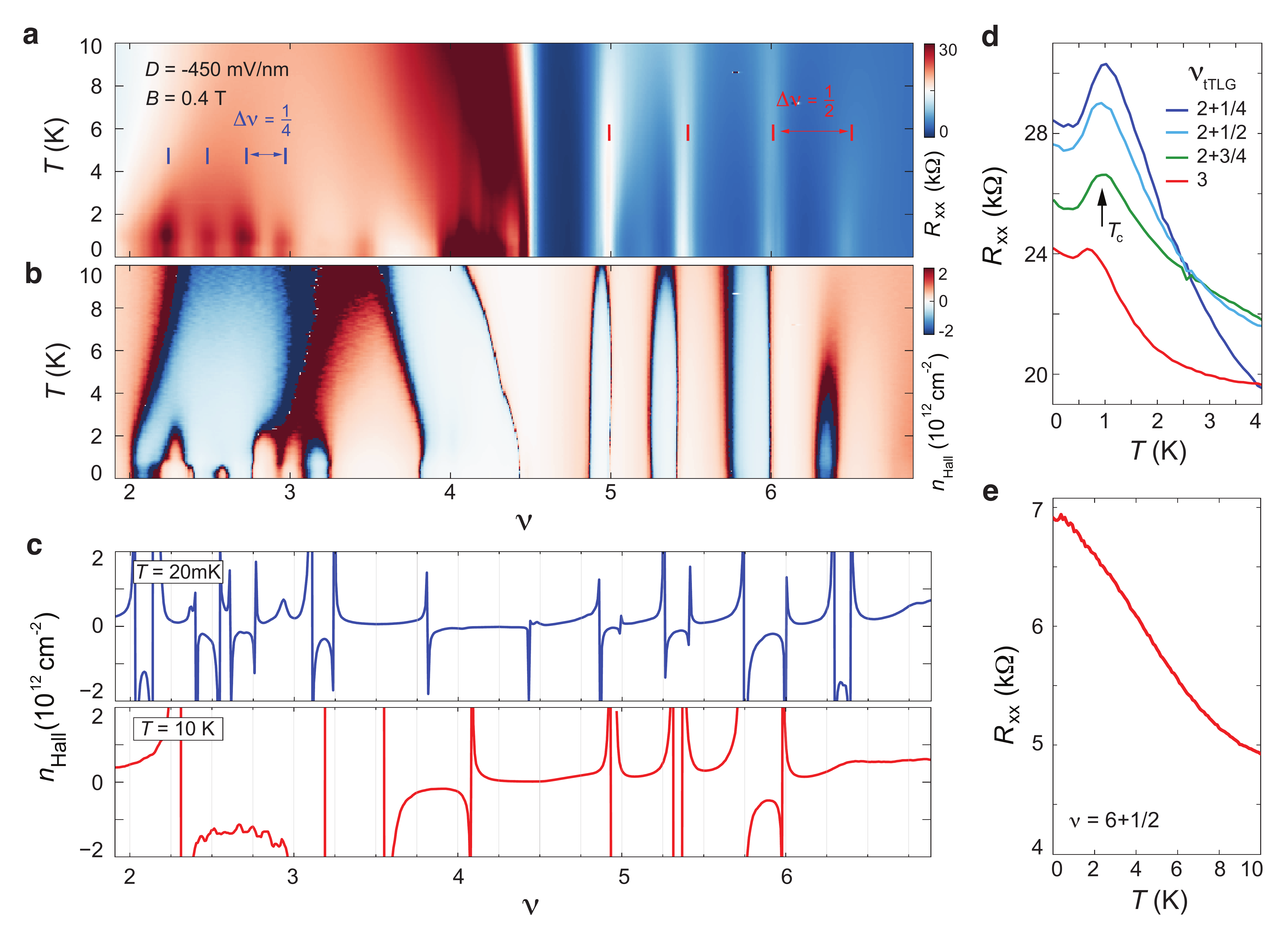}
\caption{\label{HallT} {\bf{The temperature dependence of DW states.}} (a) Longitudinal resistance \Rxx\ and (b) Hall density $n_{Hall}$ as a function of $\nu_{tTLG}$ and temperature $T$ measured at $D = -450$ mV/nm and $B = 0.4$ T. (c) $n_{Hall}$ as a function of $\nu_{tTLG}$ measured at $T=20$ mK (upper panel) and $10$ K (lower panel). (d-e) \Rxx\ as a function of temperature measured at (d) $1/4$ and (e) $1/2$ DW states. The measurement is performed with $B=0.4$ T. 
The DW states exhibit a melting transition with increasing temperature, which is evidenced by the non-monotonic temperature dependence in \Rxx\ (Panel a). The melting temperature $T_c$ is defined as the temperature where \Rxx\ is maximal. For DW states in the density range of $2 < \nu_{tTLG} < 4$, $T_c$ is shown to be $ \sim 1$ K  (Panel d). At this temperature, the vHSs also disappear (Panel b and c), providing further confirmation that the DW instability is Coulomb driven. Most importantly, the lack of hierarchy is further demonstrated by the melting temperature, which is similar between integer and fractional fillings (Panel d). Additionally, we note that a melting temperature of $\sim 1$ K is much weaker compared to the correlated insulators at integer filling of graphene moir\'e systems without proximity effect, which typically has an energy gap of $10-30$~K~\cite{Cao2018a,Cao2018b}. }}
\end{figure*}

\begin{figure*}[h]
\includegraphics[width=0.95\linewidth]{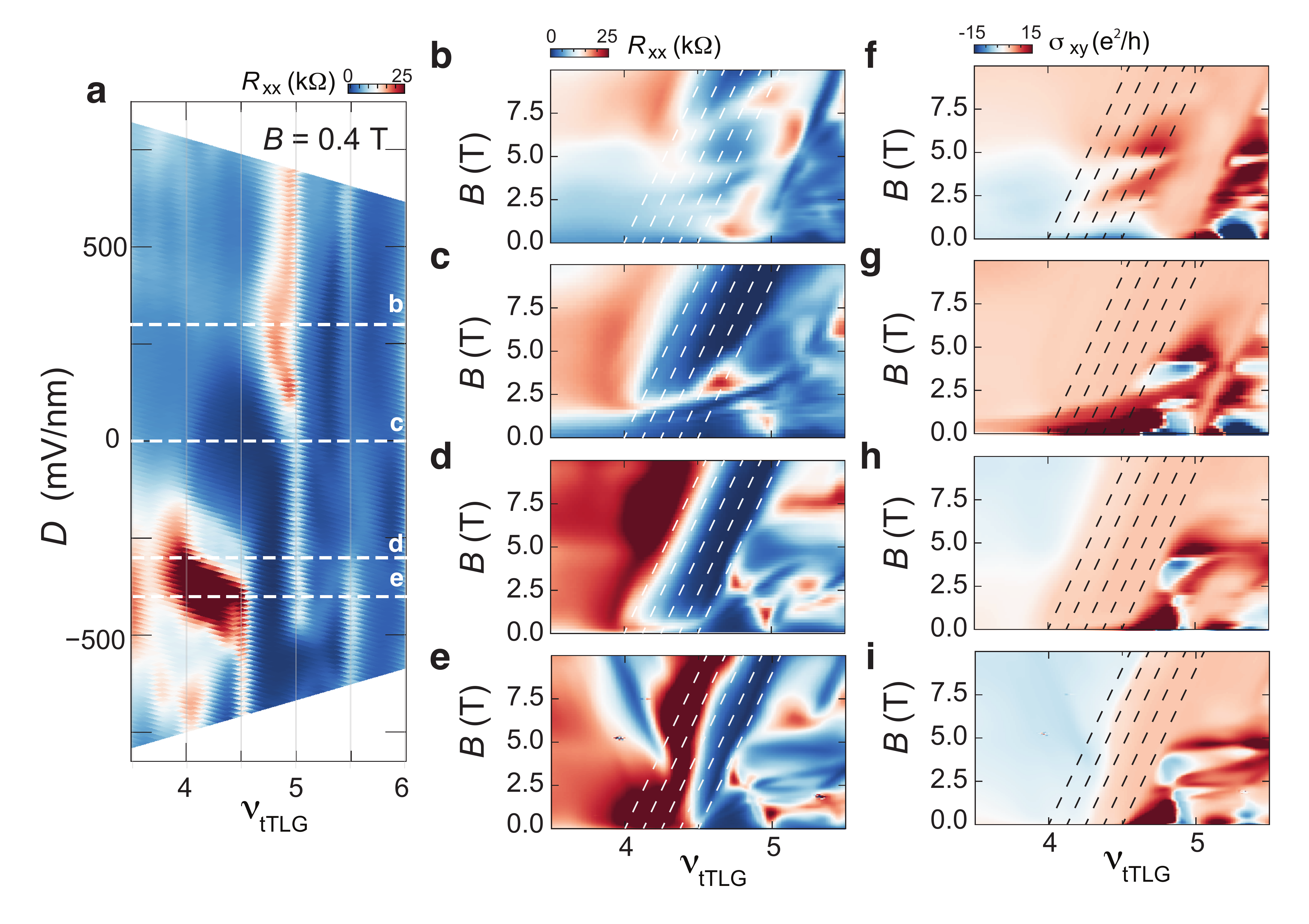}
\caption{\label{fig:layer_tran} {\bf{Influence of $D$ on incompressible states with $s=+4$.}} (a) Longitudinal resistance $R_{xx}$ as a function of displacement field $D$ and filling fraction $\nu_{tTLG}$ around $\nu_{tTLG}=+4$. (b - i) Longitudinal resistance $R_{xx}$ and Hall conductance $\sigma_{xy}$ as a function of magnetic field $B$ and filling fraction $\nu_{tTLG}$ showing incompressible states with $t=+2$. The measurement is performed at (b) (f) $D$ = 300 mV/nm,  (c) (g) $D$ = 0 mV/nm (d) (h) $D$ = -300 mV/nm, and (e) (i) $D$ = -400 mV/nm. Varying $D$ influences the stability of the incompressible state: in panel (b) the incompressible state disappears up to $B = 8$ T, pointing towards large density of state near the transition between the primary and remote bands. At the same time, the incompressible states exhibits a shift in $s$ with varying $D$, which is indicative of more complex band reconstruction between the primary and remote bands. Hall conductance measurement in panel (h) and (i) shows that the plateau shifts by $1/4$ moir\'e filling between $D=-300$ and $-400$ mV/nm, suggests that this shift could be influenced by DW instability. Although we do not fully understand this $D$ dependence, it could naturally result from the complex interaction between multiple energy bands near the transition between the flat and remote bands. Fig.~\ref{HallT}b shows that the vHS at $\nu=4.5$ and low temperature shifts back towards $\nu=4$ at $T=10$ K. This indicates that the $D$-induced shift in the Landau fan at low temperature may be Coulomb-driven. As the influence of Coulomb correlation diminishes at higher temperature, the shift disappear as well. }
\end{figure*}

\begin{figure*}[h]
{\includegraphics[width=0.85\textwidth,clip]{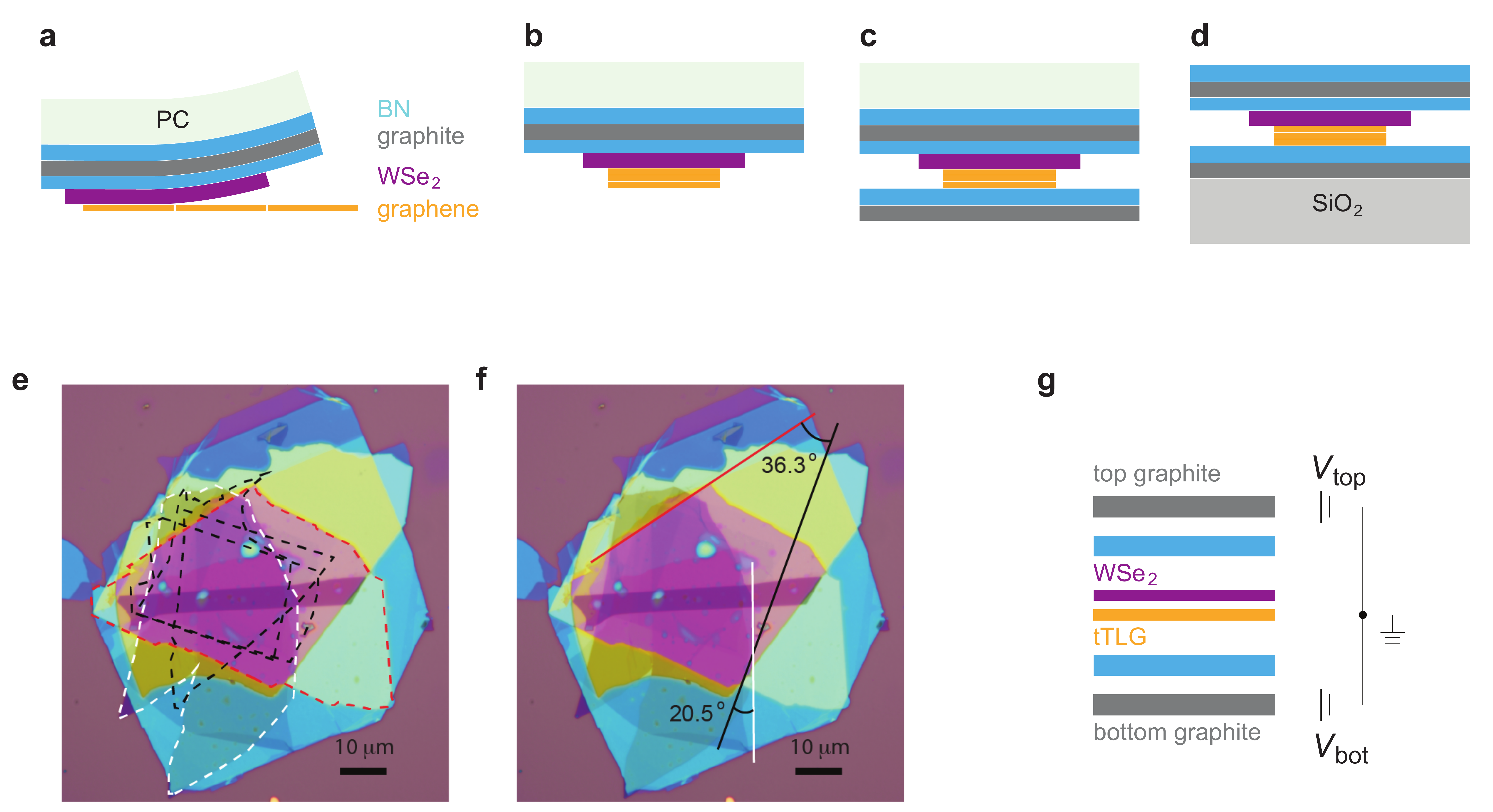}
\caption{\label{f:fab}\textbf{Fabrication of the sample.} (a-d) Schematics for stacking tTLG with cut-and-stack technique. (e) The optical image of the sample. The black dashed line, white dashed line, and red dashed line represent the boundary of graphene, \WSe, and bottom BN respectively. (f) The alignment of each crystal. The optical image shows the angle between graphene and \WSe to be 20.5 $^\circ$. and the angle between graphene and bottom BN to be 36.3 $^\circ$. (g) The schematic of the device with the gate-controlled voltage}}
\end{figure*}

\begin{figure*}[h]
{\includegraphics[width=0.75\textwidth,clip]{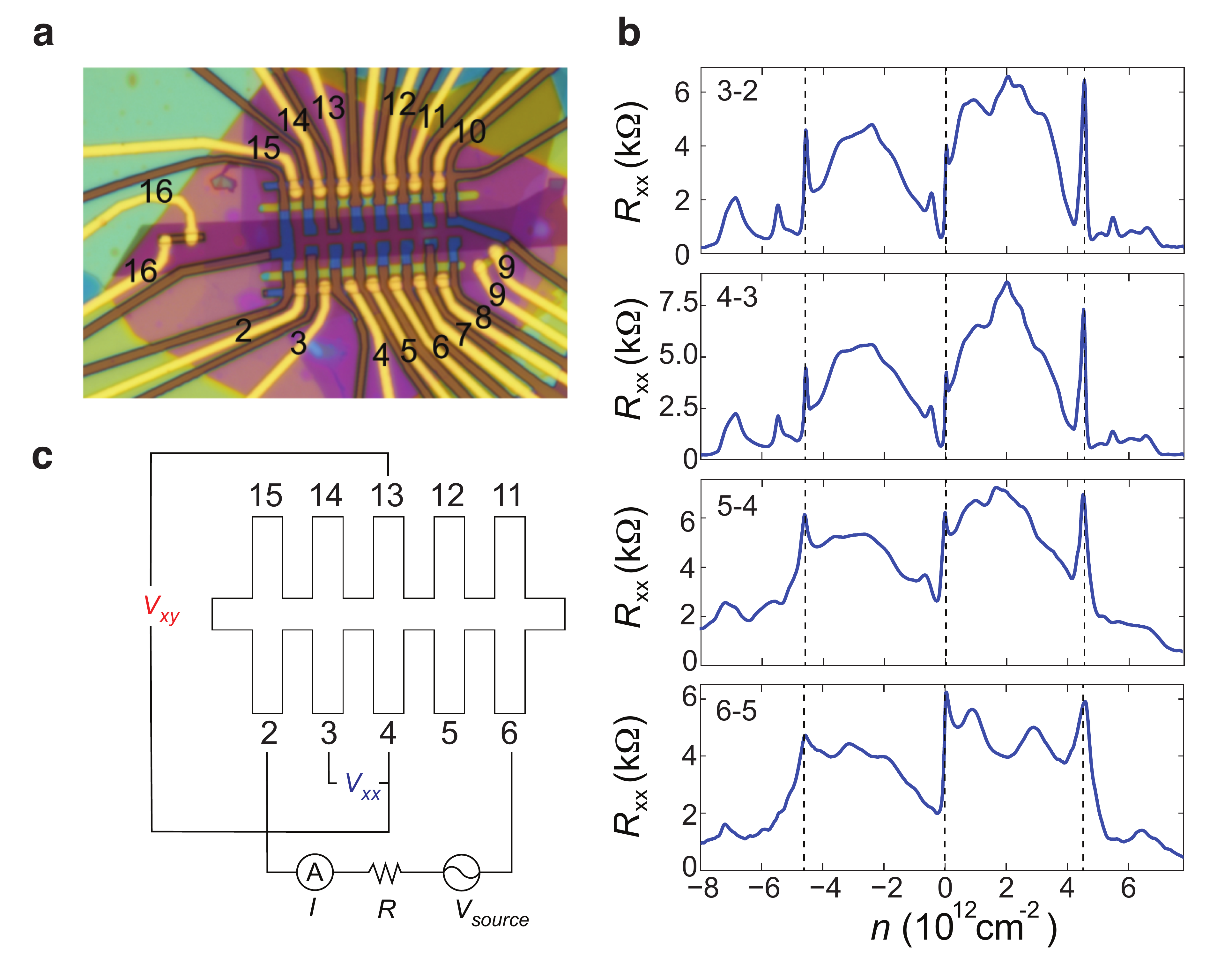}
\caption{\label{f:two-terminal}\textbf{Two-terminal measurement.}(a) an optical image of the sample with labeled lead. (b) The two-terminals measurement at different pin with $D = 0$ mV/nm. The current source and sink is placed on pin 10 and 15 respectively. Within the pin range from 2-6, the longitudinal resistances show matching peak, confirming in twist-angle homogeneity in the region.(c) the schematic of the four-terminal measurement setup for the rest of the experiment where $R_{xx} := V_{xx} / I$ and $R_{xy} := V_{xy}/I$. }}
\end{figure*}

\begin{figure*}[h]
{\includegraphics[width=0.75\textwidth,clip]{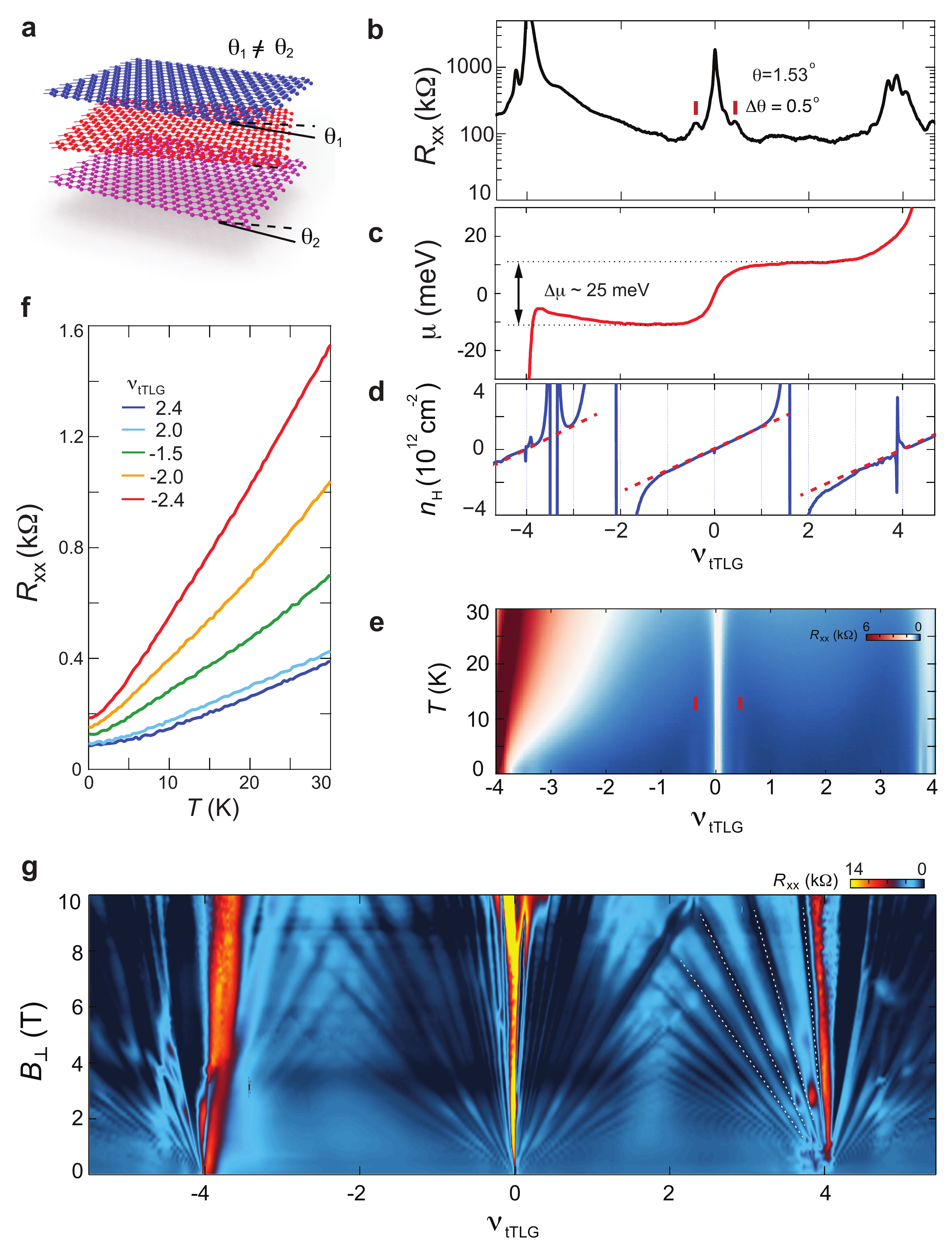}
\caption{\label{f:misalign}\textbf{Misalignment between top and bottom graphene.} In tTLG samples where top and bottom graphene layers are misaligned, transport measurement shows single particle behaviors. This type of misaligned samples do not show superconductivity down to temperature of $T = 20$ mK. (a) an optical image of the sample with labeled lead. (b) The two-terminals measurement at different pin with $D = 0$ mV/nm. The current source and sink is placed on pin 10 and 15 respectively. Within the pin range from 2-6, the longitudinal resistances show matching peak, confirming in twist-angle homogeneity in the region.(c) the schematic of the four-terminal measurement setup for the rest of the experiment where $R_{xx} := V_{xx} / I$ and $R_{xy} := V_{xy}/I$. }}
\end{figure*}

\end{widetext}
\end{document}